\documentclass[twocolumn,preprintnumbers,amsmath,amssymb]{revtex4-1}
\usepackage{graphicx} 
\usepackage{dcolumn} 
\usepackage{bm} 
\usepackage{braket} 
\usepackage[usenames,dvipsnames]{color}
\usepackage[
colorlinks=true, citecolor=blue, urlcolor=blue, linkcolor=blue,
setpagesize=false, bookmarks=false
]{hyperref}

\usepackage[english]{babel}
\addto{\captionsenglish}{}

\begin{document}


\title{
Third-harmonic generation in excitonic insulators
}
\author{Tetsuhiro Tanabe$^{1}$, Tatsuya Kaneko$^{2}$, Yukinori Ohta$^{1}$}
\affiliation{
$^1$Department of Physics, Chiba University, Chiba 263-8522, Japan\\
$^2$Department of Physics, Columbia University, New York, New York 10027, USA 
}
\date{\today}


\begin{abstract}
We study third-harmonic generation (THG) in an excitonic insulator (EI) described in a two-band correlated electron model. 
Employing the perturbative expansion with respect to the external electric field, we derive the THG susceptibility taking into account the collective dynamics of the excitonic order parameter.  
In the inversion-symmetric EI, the collective order parameter motion is activated at second order of the external field and its effects arise in THG. 
We find three peaks in the THG susceptibility at energies $\hbar \Omega = \Delta_g/3$, $\Delta_g/2$, and $\Delta_g$, where $\Delta_g$ is the band gap. 
While the THG response at $\Delta_g/3$ is caused by bare three-photon excitation of the independent particle across the band gap, the latter two peaks involve 
the motion of the order parameter activated at second order.   
The resulting resonant peaks are prominent in particular in the BCS regime but they become less significant in the BEC regime. 
We demonstrate that the resonant peak originated by the collective excitation is observable in the temperature profile of the THG intensity. 
Our study suggests that the THG measurement should be promising for detecting the excitonic collective nature of materials. 
\end{abstract}

\maketitle


\section{Introduction}

Unveiling optical properties of collective phenomena is a key issue for understanding electronic ordered states~\cite{basov2011,giannetti2016}. 
Among them, the ordered state of electron-hole pairs, the so-called excitonic insulating (EI) state~\cite{mott1961,knox1963,descloizeaux1965,keldysh1965,jerome1967,halperin1968,kunes2015}, attracts interests stimulated by recent experiments~\cite{lu2017,kogar2017,jia2020}.  
The EI states are characterized by the spontaneous band hybridization driven by the interband Coulomb interaction in narrow-gap semiconductors and semimetals, which can host ferroelectricity~\cite{batyev1980,portengen1996,batista2002,kaneko2016,kaneko2021}, magnetism~\cite{brydon2009,kaneko2012,kunes2014Co,nasu2016,yamaguchi2017,geffroy2019,nishida2019}, and topological physics~\cite{wang:2019,perfetto2020,varsano2020,sun2021TCP,liu2021,sun2021SJE}, depending on spin and orbital textures of valence and conduction bands. 
In analogy with exciton condensation, the EI is also concerned with the physics of the BCS-BEC crossover by tuning the band gap from negative (semimetal) to positive (semiconductor)~\cite{littlewood2004,bronold2006,Ihle2008,seki2011,zenker2012}.  
Recently, several transition-metal compounds, including TiSe$_2$~\cite{cercellier2007,monney2009,kogar2017,kaneko2018}, Ta$_2$NiSe$_5$~\cite{wakisaka2009,kaneko2013,*kaneko2013e,lu2017,sugimoto2018,lee2019,matsubayashi2021,fukutani2021}, and WTe$_2$~\cite{,wang2021,lee2021,jia2020},  are considered as candidates for the EIs.
In particular, the origin of the ordered state in Ta$_2$NiSe$_5$ are actively debated by the Raman and nonequilibrium pump-probe spectroscopies~\cite{mor2017,mor2018,werdehausen2018,okazaki2018,ning2020,kim2020,kim2021,volkov2021npj,ye2021,bretscher2021,suzuki2021,saha2021,bretscher2021_SciAdv,baldini2020,volkov2021}.

Dynamical properties of quantum coherent states are characterized by collective excitations. 
When the symmetry is broken spontaneously, a condensate possesses collective modes, e.g., amplitude (Higgs) mode and phase (Goldstone) mode, associated with fluctuations of an order parameter~\cite{pekker2015}.  
Recently, collective natures of materials are investigated by nonlinear optical spectroscopies.  
For example, in BCS superconductors, 
the amplitude (Higgs) mode, which is dark in linear response regime (in the long-wavelength limit), is activated by the nonlinear optical drive and the resulting resonance emerges in third-harmonic generation (THG)~\cite{tsuji2015,cea2016,tsuji2016,tsuji2020,schwarz2020,seibold2021}. 
Actually, the enhancement of the THG intensity at the resonant frequency has been observed by the terahertz pump-probe experiments~\cite{matsunaga2014,matsunaga2017,matsunaga2017PS,shimano2020,chu2020}. 

The collective excitations in the EI are also characterized by the amplitude and phase modes of the order parameter fluctuations~\cite{murakami2017,murakami2020,golez2020}. 
When an EI state is ferroelectric or breaks the spatial inversion symmetry, these two collective modes can couple to light linearly~\cite{kaneko2021}. 
However, most of the EI candidates are centrosymmetric. 
The collective modes of the inversion-symmetric EI are optically inactive in the linear response regime unless the light couples to the specific dipole~\cite{murakami2017,tanaka2018,golez2020,murakami2020}, so that we expect that the collective properties of the typical EIs strongly appear in THG, as in the superconductors. 
However, while the light-induced nonequilibrium dynamics in the EI and its candidate materials are actively investigated, the study of THG in the EI has not so far been well-developed theoretically.  

\begin{figure}[t]
    \begin{center}
    \includegraphics[bb=0 50 860 380, width=0.9\columnwidth]{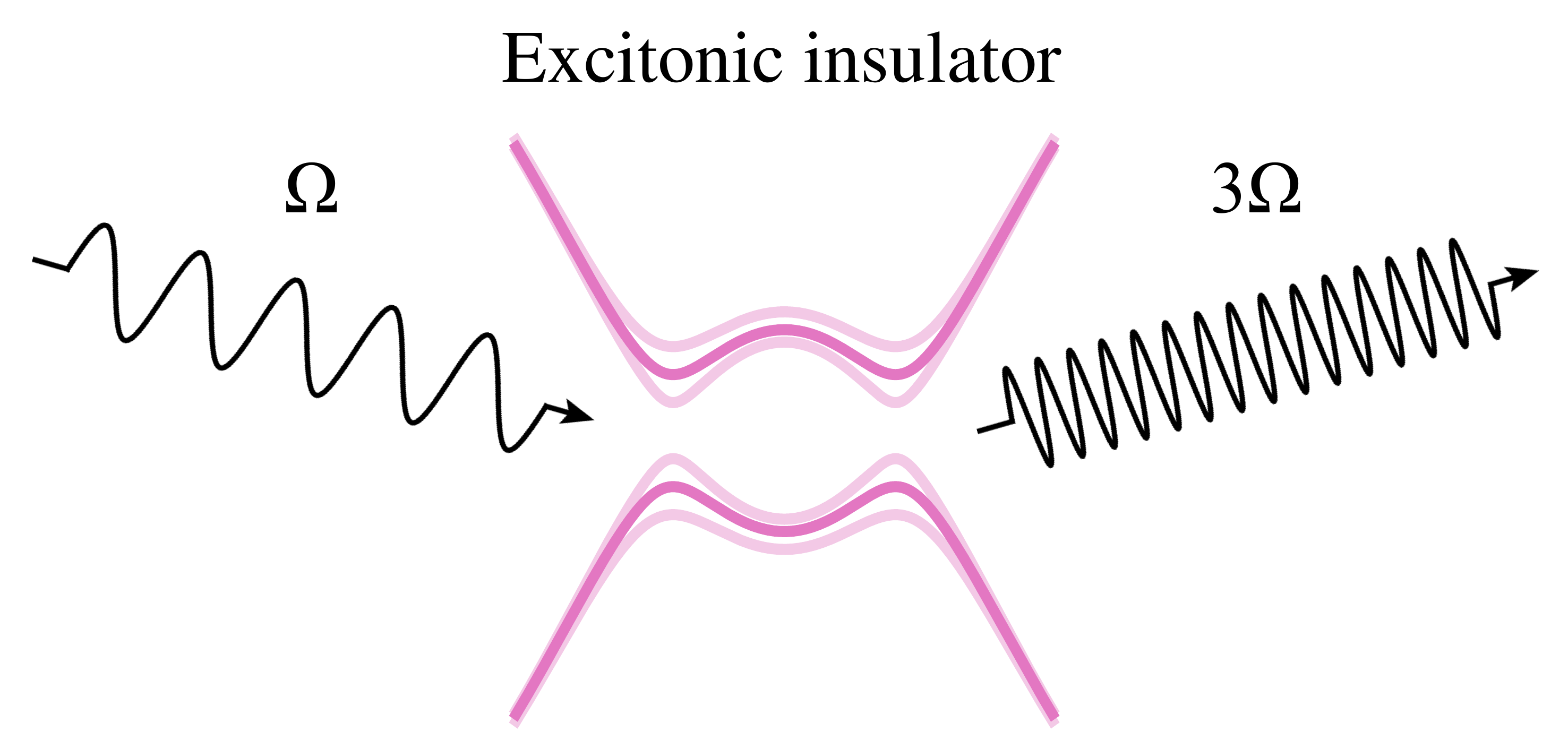}
    \end{center}
    \caption{Schematic picture of THG in the excitonic insulator.}
    \label{THGinEI}
\end{figure}

In this paper, to address this issue, we study THG in an EI described by a two-band correlated electron model (see Fig.~\ref{THGinEI}).  Employing the time-dependent mean-field theory and the perturbative expansion with respect to the external electric field, we derive the THG susceptibility taking into account the collective order parameter dynamics.   
We show that the order parameter in the inversion symmetric EI gets into motion at second order of the external field and its effects are emergent in THG. 
We find three peaks in the THG susceptibility at energies $\hbar \Omega = \Delta_g/3$, $\Delta_g/2$, and $\Delta_g$, where $\Delta_g$ is the band gap in equilibrium. 
While THG at $\Delta_g/3$ is simply originated by bare three-photon excitation of the independent particle, the latter two peaks are attributed to the motion of the order parameter activated at second order.  
The collective excitonic mode in the BCS (semimetallic) regime enhances the THG intensity  resonantly but the effect becomes less significant in the BEC (semiconducting) regime. 
From the analysis of the nonlinear response function of the excitonic order parameter, we reveal the origin of the peaks at $\Delta_g/2$, and $\Delta_g$.   
We also discuss the temperature dependence of THG and demonstrate that  the resonant peak originated by the collective motion is observable in the temperature profile of THG. 

The rest of this paper is organized as follows.  
In Sec.~\ref{sec:model} we introduce the model and time-dependent mean-field theory for the EI.  
Then, in Sec.~\ref{sec:NR}, we estimate the order parameter activated in the nonlinear regime and derive the THG susceptibility taking into account the vertex corrections.  
We show the calculated THG susceptibility in Sec.~\ref{sec:THG}. 
Discussions and summary are given in Sec.~\ref{sec:summary}.


\section{Model} \label{sec:model}

\subsection{Two-band model}
As a minimal theoretical model of the EI, we consider the spinless two-band correlated model (or extended Falicov-Kimball model)~\cite{Ihle2008,seki2011,zenker2012,kaneko2013_EFKM,ejima2014,seki2014,hamada2017,kadosawa2020}. 
The Hamiltonian takes the form 
\begin{align}
    \hat{H} = \hat{H}_0 + \hat{H}_{\rm int},
\end{align}
with 
\begin{align}
    &\hat{H}_0
        = -\sum_{\langle i,j\rangle}\sum_{\alpha}
            \left(
                t_{\alpha} \hat{c}_{i,\alpha}^{\dagger}\hat{c}_{j,\alpha}
                +{\rm H.c.}
            \right)
            + \sum_{j,\alpha} \Delta_{\alpha} \hat{c}_{j,\alpha}^{\dagger}\hat{c}_{j,\alpha}, 
            \\
        &\hat{H}_{\rm int} = U \sum_j 
            \hat{c}_{j,0}^{\dagger} \hat{c}_{j,0}
            \hat{c}_{j,1}^{\dagger} \hat{c}_{j,1}, 
\end{align}
where $\hat{c}_{j,\alpha}$ ($\hat{c}_{j,\alpha}^{\dag}$) is the annihilation  (creation) operator of an electron at site $j$ on orbital $\alpha$ ($=0,1$), and $\langle i,j \rangle$ indicates a pair of nearest-neighbor sites. 
$t_{\alpha}$, $\Delta_{\alpha}$, and $U$ are the hopping integral, energy level of the orbital $\alpha$, and interorbital repulsive interaction, respectively. 
Here we focus on the half-filled case $n_0 + n_1 = \langle \hat{c}_{j,0}^{\dagger} \hat{c}_{j,0} \rangle + \langle \hat{c}_{j,1}^{\dagger} \hat{c}_{j,1} \rangle=1$ and consider the model defined on the two-dimensional (2D) square lattice ($d=2$). 
The free electron part in the momentum ($\bm{k}$) space is given by
\begin{align}
    &\hat{H}_0
             = \sum_{\bm{k},\alpha}
            \varepsilon_{\alpha}(\bm{k})
            \hat{c}_{\bm{k},\alpha}^{\dagger}\hat{c}_{\bm{k},\alpha}, 
            \\
    &\varepsilon_{\alpha}(\bm{k})
        \equiv -2t_{\alpha} \left[ \cos{\left(k_x a\right)} +\cos{\left(k_y a\right)} \right] + \Delta_{\alpha}, 
\end{align}
where we use the Fourier transformation $\hat{c}_{j,\alpha} = \frac{1}{\sqrt{N}} \sum_{\bm{k}} e^{i\bm{k}\cdot\bm{R}_j} \hat{c}_{\bm{k},\alpha}$ ($N$ is the number of lattice site), and $a$ is the lattice constant. 
We take the particle-hole symmetric band structure with $t_0 = -t_1$ (direct-gap type) and assume $\Delta_0 + \Delta_1 = -U$ in order to set the Fermi energy to zero. 

The external field $\bm{A}(t)$ is introduced by the Peierls substitution~\cite{tanabe2018,fujiuchi2019}, and we use the time-dependent Hamiltonian $\hat{H}(t) = \hat{H}_0(t) + \hat{H}_{\rm int}$, with 
\begin{align}
    \hat{H}_0(t)
        = \sum_{\bm{k},\alpha}
            \varepsilon_{\alpha}\left(\bm{k}+\hbar^{-1}e\bm{A}(t)\right)
            \hat{c}_{\bm{k},\alpha}^{\dagger}\hat{c}_{\bm{k},\alpha}, 
\end{align}
where $e$ ($>0$) is the elementary charge and $\hbar$ is the Plank constant. 
In this paper we use the monochromatic continuous-wave $\bm{A}(t)=\bm{A}(\Omega) e^{-i\Omega t} + {\rm c.c.}$ unless otherwise noted. 
We assume that the interorbital dipole coupling $\bm{d}$~\cite{murakami2017,tanaka2018,golez2020,murakami2020} is zero for simplicity because it depends on the parities of the two orbitals.

\subsection{Mean-field theory}
In this paper we employ the time-dependent mean-field (tdMF) theory~\cite{murakami2017,murakami2020,golez2020}. 
We define the mean values of the diagonal and off-diagonal densities as 
\begin{align}
    n_{\alpha}(t) = \left\langle \hat{c}_{j,\alpha}^{\dagger}(t) \hat{c}_{j,\alpha}(t) \right\rangle, \;\;\;
    \phi(t) = \left\langle \hat{c}_{j,0}^{\dagger}(t) \hat{c}_{j,1}(t) \right\rangle, 
\end{align}
respectively, where the off-diagonal component $\phi(t)$ corresponds to the order parameter of the EI in our two-band model. 
Then, the MF Hamiltonian is given by
\begin{align}
    \hat{H}(t) \longrightarrow \hat{H}_{\mathrm{MF}}(t) =
        \sum_{\bm{k}}\sum_{\alpha,\alpha'}h_{\alpha\alpha'}^A(\bm{k},t)
        \hat{c}_{\bm{k},\alpha}^{\dagger} \hat{c}_{\bm{k},\alpha'},
\end{align}
with 
\begin{align}
      &h^A(\bm{k},t)
        = h\left(\bm{k}+\hbar^{-1}e\bm{A}(t),\,t\right) , 
        \\
       &h(\bm{k},t)
        = \left[
            \begin{array}{cc}
              \varepsilon_0(\bm{k})+Un_1(t) & -U\phi^*(t) \\
              -U\phi(t) & \varepsilon_1(\bm{k})+Un_0(t) \\
            \end{array}
        \right] , 
\end{align}
where $h^A(\bm{k},t)$ is the matrix on the basis $\hat{\Psi}^{\dagger}_{\bm{k}} = [\hat{c}_{\bm{k},0}^{\dagger}  \, \hat{c}_{\bm{k},1}^{\dagger} ]$. 

In the pseudospin representation, the Hamiltonian $h^A(\bm{k},t)$ is described by
\begin{align}
&    h^A(\bm{k},t)=\frac{1}{2}\bm{B}(\bm{k},t)\cdot\bm{\sigma} + \frac{1}{2}B_0(\bm{k},t) \sigma_0, 
\end{align}
where $\sigma_0$ and $\sigma_a$ ($a=x,y,z$) are the identity and Pauli matrices, respectively, 
\begin{align}
    B_x(\bm{k},t) =& -2U\,{\mathrm{Re}}\,\phi(t), \label{pseudomagfield_1} \\
    B_y(\bm{k},t) =& -2U\,{\mathrm{Im}}\,\phi(t), \label{pseudomagfield_2} \\
    B_z(\bm{k},t) =& \; \varepsilon_0(\bm{k}+\hbar^{-1}e\bm{A}(t))-\varepsilon_1(\bm{k}+\hbar^{-1}e\bm{A}(t))
    \notag \\
        &-U \left(n_0(t)-n_1(t)\right) \label{pseudomagfield_3}, 
\end{align}
and $B_0(\bm{k},t)=0$ since we assume $t_0=-t_1$ and $\Delta_0 + \Delta_1 = -U$. 
Note that $B_0(\bm{k},t) = \varepsilon_0(\bm{k}+\hbar^{-1}e\bm{A}(t))+\varepsilon_1(\bm{k}+\hbar^{-1}e\bm{A}(t))+U$ when $t_0\ne-t_1$ and $\Delta_0 + \Delta_1 \ne -U$ at half-filling.   
In  the pseudospin representation, the MF parameter is given by 
\begin{align}
     \phi_a(t) = \frac{1}{2N} \sum_{\bm{k}} \left\langle \hat{\Psi}_{\bm{k}}^{\dagger} (t) \sigma_a\hat{\Psi}_{\bm{k}}(t)  \right\rangle, 
\end{align}
which composes the vector 
\begin{align}
    \bm{\phi}(t)
    =\left[
        \begin{array}{c}
            \phi_x(t) \\
            \phi_y(t) \\
            \phi_z(t)
        \end{array}
    \right]
    =\left[
        \begin{array}{c}
            \mathrm{Re}\,\phi(t) \\
            \mathrm{Im}\,\phi(t) \\
            \left[n_0(t)-n_1(t)\right]/2
        \end{array}
    \right]. 
\end{align}
Then, the vector $\bm{B}(\bm{k},t)$ is given by 
\begin{align}
    \bm{B}(\bm{k},t) = - 2 U \bm{\phi}(t) +2  \xi(\bm{k}+\hbar^{-1}e\bm{A}(t)) \bm{e}_z, 
\end{align}
where we define 
\begin{align}
            \xi(\bm{k}) \equiv \frac{\varepsilon_0(\bm{k})-\varepsilon_1(\bm{k})}{2}.
\end{align}
In the tdMF theory, the time-dependent current is determined by 
\begin{align}
    J_{\mu}(t)
        = -e\sum_{\bm{k}}v_{\mu}\left(\bm{k}+\hbar^{-1}e\bm{A}(t)\right)
            \left\langle \hat{\Psi}^{\dagger}_{\bm{k}}(t) \sigma_z \hat{\Psi}_{\bm{k}}(t) \right\rangle,
\end{align}
where 
\begin{align}
&v_{\mu}(\bm{k})
        \equiv \frac{1}{\hbar}
            \frac{\partial \xi({\bm{k}})}{\partial k_{\mu}}.
\end{align}
Note that, when $t_0 \ne -t_1$, we need to include the $\sigma_0$ component in the current. 
When we perform the real-time simulations, we solve the equation of motion $\partial_t \bm{S}(\bm{k},t)=  \hbar^{-1} \bm{B}(\bm{k},t)\times \bm{S}(\bm{k},t)$ for $S_a(\bm{k},t) \equiv \langle \hat{\Psi}_{\bm{k}}^{\dagger} (t) \sigma_a\hat{\Psi}_{\bm{k}}(t)\rangle/2$ with updating the MF parameter $\phi_a(t)$ simultaneously. 
In this paper we expand the nonequilibrium quantities and Green's function with respect to the external field $\bm{A}(t)$ and estimate the photocurrent for THG. 

In equilibrium [$B_a(\bm{k})=B_a(\bm{k},t=-\infty)$] we have the eigenenergy 
\begin{align}
    E_{\pm}(\bm{k}) = \hbar \omega_{\pm}(\bm{k}) = \pm\frac{1}{2}\left|\bm{B}(\bm{k})\right|+\frac{1}{2}B_{0}(\bm{k}) , 
\end{align}
and the MF parameter is determined by 
\begin{align}
     \phi_a = \frac{1}{N} \sum_{\bm{k}} \frac{B_a(\bm{k})}{2\left|\bm{B}(\bm{k})\right|}
     \left[ f(E_+(\bm{k}))-f(E_-(\bm{k})) \right], 
\end{align}
where $f(E)$ is the Fermi distribution function. 
We solve this equation self-consistently and determine the MF parameters in equilibrium. 
The bare lesser ($<$) and retarded/advanced ($R/A$) Green's functions are given by 
\begin{align}
    &G^{0,<} (\bm{k},t)
        = i\sum_{\nu=\pm} f(E_{\nu}(\bm{k}))b_{\nu}(\bm{k})e^{-i\omega_{\nu}(\bm{k})t}, \\
    &G^{0,R/A} (\bm{k},t)
        =  \mp i\theta(\pm t)\sum_{\nu=\pm} b_{\nu}(\bm{k}) e^{-i\omega_{\nu}(\bm{k})t}, 
\end{align}
respectively, where 
\begin{align}
    b_{\nu}(\bm{k})
        =\frac{1}{2} \left[ \sigma_0 + \nu \frac{\bm{B}(\bm{k}) }{\left|\bm{B}(\bm{k})\right|}\cdot \bm{\sigma} \right].
\end{align}
In the following we also use the Fourier transformed Green's function $ G^{0} (\bm{k},\omega) = \int dt G^{0} (\bm{k},t) e^{i\omega t}$.


\section{Nonlinear Responses} \label{sec:NR}

\subsection{Perturbative expansion}

Using the nonequilibrium Green's function $G(\bm{k},t,t')$ under the applied external field $\bm{A}(t)$ (see Appendix~\ref{App:GF}), the MF parameter and current are given by 
\begin{align}
    &\phi_{a}(t)
        = -i\frac{1}{2N} \sum_{\bm{k}} {\mathrm{tr}}\left[\sigma_a G^{<}(\bm{k},t,t)\right], 
        \label{eq:OPGF}
        \\
         &J_{\mu}(t)
        = ie\sum_{\bm{k}}v_{\mu}\left(\bm{k}+\hbar^{-1}e\bm{A}(t)\right)
            \mathrm{tr}\left[\sigma_z G^{<}(\bm{k},t,t)\right] , 
            \label{eq:JGF}
\end{align}
respectively. 
In this section we expand the Green's function (and velocity) with respect to the external field $\bm{A}(t)$ and derive the order parameter and current induced in the nonlinear regime. 

With respect to $\bm{A}(t)$, we expand a quantity $X$ as 
\begin{align}
    X(\bm{A})&=\sum_{n=0}^{\infty} X^{(n)}(\bm{A})
    =\sum_{n=0}^{\infty}\frac{1}{n!} \delta^n X (\bm{A}), 
\end{align}
where $X^{(n)}(\bm{A})= \delta^n X (\bm{A})/n! =  \mathcal{O}(A^n)$. 
In this notation, the $n$th order variation of the Hamiltonian $h^A(\bm{k},t)$ is given by 
\begin{align}
    &\delta^n h^A(\bm{k},t)
        = -U \delta^n \bm{\phi}(t)\cdot \bm{\sigma}
        \\
          &  + \left( \frac{e}{\hbar} \right)^n \sum_{\mu_1,\cdots,\mu_n}
            \xi_{\mu_1\mu_2\cdots \mu_n}(\bm{k})
            A_{\mu_1}(t)A_{\mu_2}(t) \cdots A_{\mu_n}(t)
            \,\sigma_z , 
            \notag
\end{align}
where 
\begin{align}
     \xi_{\mu_1\cdots\mu_n}(\bm{k})
        \equiv
            \frac{\partial^n \xi(\bm{k})}{\partial k_{\mu_1}\partial k_{\mu_2}\cdots\partial k_{\mu_n}}. 
\end{align}

We expand the Green's function $G$ with respect to the deviation from equilibrium $\delta^n\mathcal{H}$ given by $\delta^n h^A(\bm{k},t)$. 
The details of the nonequilibrium Green's function and its expansion are summarized in Appendix~\ref{App:GF}. 
Expanding the Green's function up to the third order, we have 
\begin{align}
    \delta G
        &= G^0 * \delta\mathcal{H} * G^0, 
        \label{eq:GF1st}
          \\
    \delta^2 G
        &= 2 G^0 * \delta\mathcal{H} * G^0 * \delta\mathcal{H} * G^0
         \nonumber\\
        & + G^0 * \delta^2\mathcal{H} * G^0, 
        \label{eq:GF2nd}
        \\
    \delta^3 G
        &= 6 G^0 * \delta\mathcal{H} * G^0
           * \delta\mathcal{H} * G^0 * \delta\mathcal{H} * G^0
               \nonumber\\
        &+ 3 G^0 * \delta\mathcal{H} * G^0 
           * \delta^2\mathcal{H} * G^0
            \nonumber\\               
        &+ 3 G^0 * \delta^2 \mathcal{H} * G^0
            * \delta\mathcal{H} * G^0
            \nonumber\\
        & + G^0 * \delta^3 \mathcal{H} * G^0, 
        \label{eq:GF3rd}
\end{align}
where $X*Y$ indicates the product including the time-integration $\int dt_1 X^{\zeta \zeta_1}(t,t_1)Y^{\zeta_1 \zeta'}(t_1,t')$ (see details in Appendix~\ref{App:GF})~\cite{aoki2014}. 

In the following, we estimate the MF parameter $\delta^2 \phi_a$ at second order and then derive the nonlinear current $\delta^3 J_{\mu}$ for THG involving the collective dynamics of the order parameter (i.e., vertex correction).

\subsection{Order parameter}

First, we derive the order parameter away from equilibrium by expanding the Green's function. 
Because of the symmetry under inversion, e.g., $\xi_{\mu}(-\bm{k})= -\xi_{\mu}(\bm{k})$, 
the MF parameters at odd order [$\delta \bm{\phi}(t)$, $\delta^3 \bm{\phi}(t)$, $\cdots$] vanish (see Appendix \ref{MFodd}). 
Hence, the lowest order of the activated order parameter is of the second order; 

\begin{align}
    \delta^2 \phi_{a}(t)
        = -i\frac{1}{2N} \sum_{\bm{k}} {\mathrm{tr}}\left[\sigma_a \delta^2 G^{<}(\bm{k},t,t)\right]. 
        \label{eq:OP2nd}
\end{align}
Under the monochromatic field $\bm{A}(t)=\bm{A}(\Omega) e^{-i\Omega t} + \bm{A}(-\Omega) e^{i\Omega t}$, the MF parameter at second order is characterized by $\delta^2 \phi_{a}(t)= \delta^2 \phi_{a}(2\Omega)e^{-2i\Omega t} + \delta^2 \phi_{a}(-2\Omega)e^{2i\Omega t} + \delta^2 \phi_{a}(0)$. 
While $\delta^2 \phi_{a}(0)$ [$\propto A(\Omega) A(-\Omega)$] can be nonzero, it does not contribute to THG given by $\delta^3 J(3\Omega)$ [$\propto A(\Omega)^3$]. 
Here, we consider $\delta^2 \phi_{a}(2\Omega)$ [$\propto A(\Omega)^2$] because THG originated from the dynamical order parameter (vertex correction) is described by $\delta^3 J^{\rm vc}(3\Omega)\propto \delta^2 \phi_a(2\Omega) A(\Omega)$.
Combining Eqs.~(\ref{eq:GF2nd}) and (\ref{eq:OP2nd}), $\delta^2 \phi_{a}(2\Omega)$ is given by 
\begin{widetext}
\begin{align}
    \delta^2 \phi_{a}(2\Omega)
        =& -i\frac{1}{2N} \sum_{\bm{k}}   \frac{2}{\hbar^2}\int \frac{d\omega}{2\pi}
                        {\mathrm{tr}}\left[\sigma_a G^{0}(\bm{k},\omega+2\Omega) \delta h^A(\bm{k},\Omega) G^{0}(\bm{k},\omega+\Omega)  \delta h^A(\bm{k},\Omega)G^{0}(\bm{k},\omega)\right]^{<} 
                        \notag\\
& -i\frac{1}{2N}\sum_{\bm{k}}   \frac{1}{\hbar} \int \frac{d\omega}{2\pi}
                        {\mathrm{tr}}\left[\sigma_a  G^{0}(\bm{k},\omega+2\Omega)  \delta^2 h^A(\bm{k},2\Omega)G^{0} (\bm{k},\omega)\right]^{<}      , 
\label{eq:OP2ndSCE}                   
\end{align}
where $[\cdots]^{<}$ indicates the lesser component following the Langreth's rule, e.g., $[XY]^{<}=X^{R}Y^{<}+X^{<}Y^A$ (see details in Appendix~\ref{App:GF}). 
While we can integrate the Green's functions over $\omega$ as summarized in Appendix~\ref{App:integw}, we retain the $\omega$ integral with the Green's functions for the compact notation.  
Since $\delta \bm{\phi}(t)=0$ (see Appendix \ref{MFodd}), we have 
\begin{align}
    &\delta h^A(\bm{k},\Omega)
        =  \frac{e}{\hbar}  \sum_{\mu_1}
            \xi_{\mu_1}(\bm{k}) A_{\mu_1}(\Omega) \,\sigma_z , 
\end{align}
but $\delta^2 \bm{\phi}(2\Omega)$ can be nonzero and 
\begin{align}
    \delta^2 h^A(\bm{k},2\Omega)
        =  - U \delta^2 \bm{\phi}(2\Omega)\cdot \bm{\sigma}
        +  \left( \frac{e}{\hbar} \right)^2 \sum_{\mu_1,\mu_2}
            \xi_{\mu_1\mu_2}(\bm{k})
             A_{\mu_1}(\Omega)A_{\mu_2}(\Omega) \,\sigma_z   . 
\end{align}
Equation~(\ref{eq:OP2ndSCE}) corresponds to the self-consistent equation of $\delta^2 \bm{\phi}(2\Omega)$ because $\delta^2 h^A(\bm{k},2\Omega)$ in the right-hand side of Eq.~(\ref{eq:OP2ndSCE}) includes $\delta^2 \bm{\phi}(2\Omega)$. 

\begin{figure*}[!t]
\begin{center}
\includegraphics[bb=0 0 470 60,width=0.75\columnwidth]{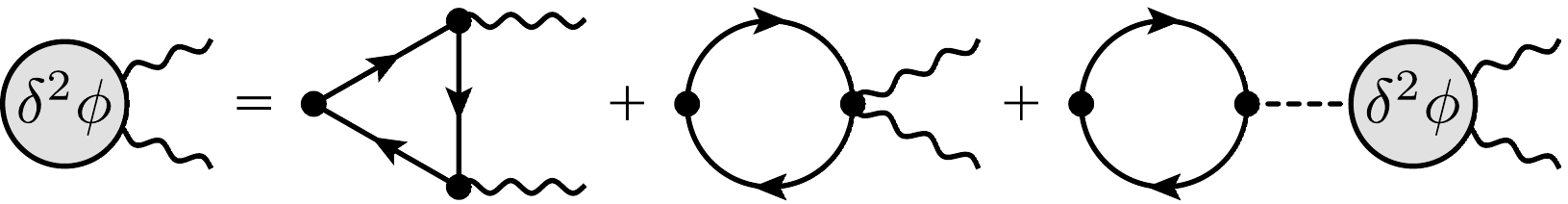}
\caption{The diagrammatic representation of Eq.(\ref{eq:OP2ndSCE_2}). The solid (with arrow), wavy, and dashed lines indicate the bare Green's function $G^0$, external field $\bm{A}$, and interaction $U$, respectively.}
\label{fig:diagOP}
\end{center}
\end{figure*}

Introducing the bare susceptibilities for coupling between the order parameter $\phi$ and the external field $A_{\mu}(\Omega)$,
\begin{align}
\bm{\chi}^{0;\phi\xi\xi}_{\mu_1\mu_2}(2\Omega;\Omega,\Omega)  &=  -i \frac{1}{\hbar^2} \frac{1}{N} \sum_{\bm{k}} \int \frac{d\omega}{2\pi}
{\mathrm{tr}}\left[\bm{\sigma} G^{0}(\bm{k},\omega+2\Omega) \sigma_z G^{0}(\bm{k},\omega+\Omega)  \sigma_z G^{0}(\bm{k},\omega)\right]^{<} \xi_{\mu_1} (\bm{k}) \xi_{\mu_2}(\bm{k}), 
\label{eq:chi_pxx}
 \\
\bm{\chi}^{0;\phi\xi}_{\mu_1\mu_2}(2\Omega;2\Omega)  &= -i \frac{1}{2\hbar} \frac{1}{N} \sum_{\bm{k}}    \int \frac{d\omega}{2\pi}
{\mathrm{tr}}\left[ \bm{\sigma} G^{0}(\bm{k},\omega+2\Omega) \sigma_z  G^{0} (\bm{k},\omega)\right]^{<} \xi_{\mu_1\mu_2}(\bm{k}), 
\label{eq:chi_px}
\end{align}
and the bare $\phi$-$\phi$ susceptibility ($3\times 3$ matrix)
\begin{align}
\left[ \tilde{\chi}^{0;\phi\phi}(2\Omega)\right]_{ab} = -i \frac{1}{2\hbar} \frac{1}{N} \sum_{\bm{k}}  \int \frac{d\omega}{2\pi}
{\mathrm{tr}}\left[\sigma_a G^{0}(\bm{k},\omega+2\Omega) \sigma_b G^{0}(\bm{k},\omega) \right]^{<} , 
\label{eq:sus_phiphi}
\end{align}
the self-consistent Eq. (\ref{eq:OP2ndSCE}) becomes
\begin{align}
\delta^2 \! \bm{\phi} (2\Omega) 
= \left( \frac{e}{\hbar}\right)^2 \! \sum_{\mu_1,\mu_2} \! \bm{\chi}^{0;\phi\xi\xi}_{\mu_1\mu_2}(2\Omega;\Omega,\Omega) A_{\mu_1}(\Omega) A_{\mu_2}(\Omega)
+\left( \frac{e}{\hbar}\right)^2 \! \sum_{\mu_1,\mu_2} \! \bm{\chi}^{0;\phi\xi}_{\mu_1\mu_2}(2\Omega;2\Omega) A_{\mu_1}(\Omega) A_{\mu_2}(\Omega)
-U \tilde{\chi}^{0;\phi\phi}(2\Omega)\delta^2 \! \bm{\phi} (2\Omega) .
\label{eq:OP2ndSCE_2}
\end{align}
This equation may be described by the diagrams in Fig.~\ref{fig:diagOP}~\cite{tsuji2015}. 
Then we obtain the solution 
\begin{align}
\delta^2 \bm{\phi} (2\Omega) =   \left( \frac{e}{\hbar}\right)^2 \sum_{\mu_1, \mu_2} \frac{\bm{\chi}^{0;\phi\xi\xi}_{\mu_1\mu_2}(2\Omega;\Omega,\Omega)}{1+U\tilde{\chi}^{0;\phi\phi}(2\Omega)} A_{\mu_1}(\Omega) A_{\mu_2}(\Omega)
+  \left( \frac{e}{\hbar}\right)^2 \sum_{\mu_1, \mu_2} \frac{\bm{\chi}^{0;\phi\xi}_{\mu_1\mu_2}(2\Omega;2\Omega)}{1+U\tilde{\chi}^{0;\phi\phi}(2\Omega)} A_{\mu_1}(\Omega) A_{\mu_2}(\Omega), 
\end{align}
\end{widetext} 
indicating that the order parameter is activated at second order of $\bm{A}$ when the corrected susceptibility is nonzero. 
For later convenience we express the above relation as 
\begin{align}
\delta^2 \phi_a (2\Omega) = \left( \frac{e}{\hbar}\right)^2\sum_{\mu_1, \mu_2} \Gamma_{\mu_1\mu_2}^a(2\Omega; \Omega,\Omega) A_{\mu_1}(\Omega) A_{\mu_2}(\Omega),
\label{eq:OP2ndGamma}
\end{align}
with 
\begin{align}
\bm{\Gamma}_{\mu_1\mu_2}(2\Omega; \Omega,\Omega) 
&=\frac{\bm{\chi}^{0;\phi\xi\xi}_{\mu_1\mu_2}(2\Omega;\Omega,\Omega)}{1+U\tilde{\chi}^{0;\phi\phi}(2\Omega)}
+\frac{\bm{\chi}^{0;\phi\xi}_{\mu_1\mu_2}(2\Omega;2\Omega)}{1+U\tilde{\chi}^{0;\phi\phi}(2\Omega)}. 
\end{align}

\begin{figure*}[!t]
\begin{center}
\includegraphics[bb=0 0 468 198,width=1.89\columnwidth]{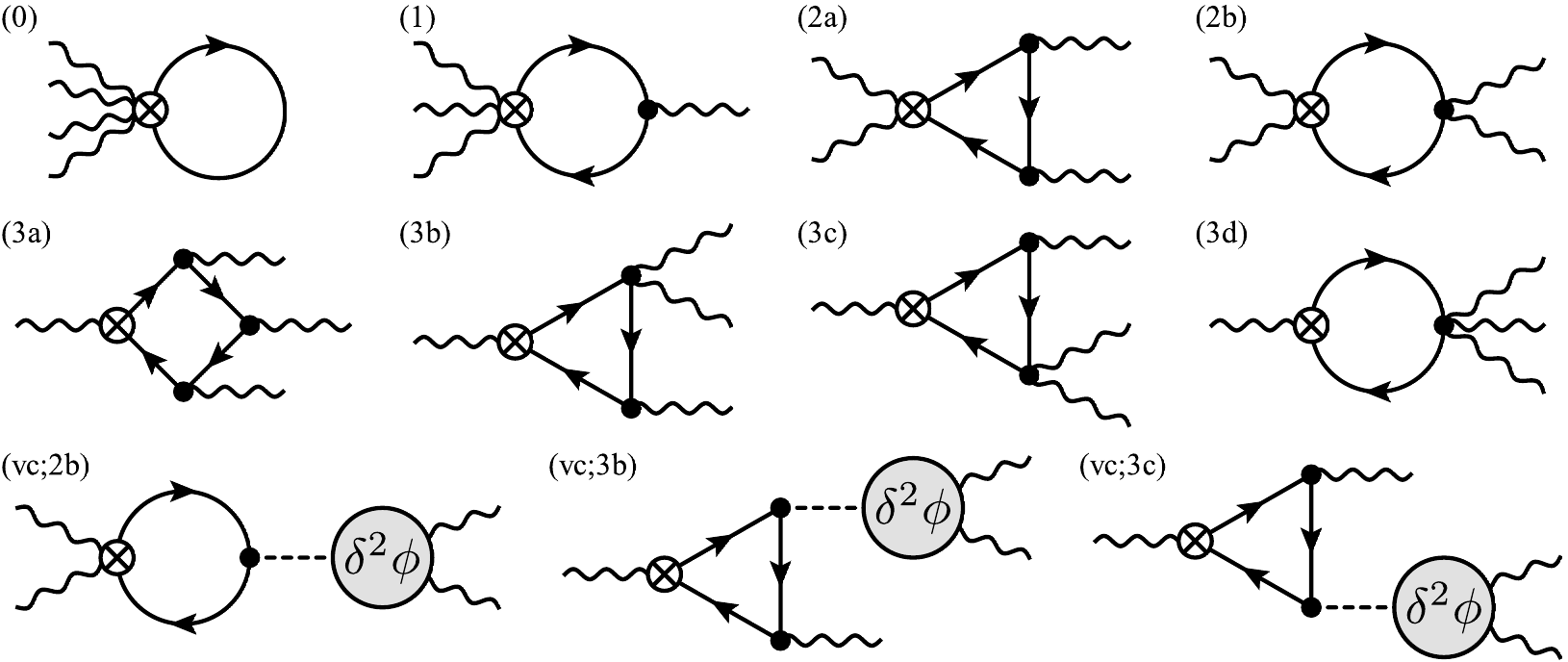}
\caption{The diagrams for the third-order photocurrent. The solid (with arrow), wavy, and dashed lines indicate the bare Green's function $G^0$, external field $\bm{A}$, and interaction $U$, respectively. The cross indicates the one-photon output.}
\label{fig:diagJ}
\end{center}
\end{figure*}

\subsection{Current}

Next, we derive the nonlinear current involving the dynamics of the order parameter. 
Expanding $J_{\mu}(t)$ in Eq.~(\ref{eq:JGF}), the current at $n$th order is given by 

\begin{align}
    \delta^n \! J_{\mu} (t) 
        \!=\!  ie \! \sum_{\bm{k}} \sum_{m=0}^n \! \binom{n}{m} \delta^{n-m}v^A_{\mu}(\bm{k},t)
        {\mathrm{tr}} \left[
                \sigma_z \delta^m G^<(\bm{k},t,t)
            \right],
            \label{eq:Jnth}
\end{align}
where $ \binom{n}{m}$ is binomial coefficient and 
\begin{align}
    \delta^{n}v^A_{\mu}(\bm{k},t)
         = \frac{e^{n}}{\hbar^{n+1}}
        \!  \sum_{\mu_1,\cdots,\mu_{n}} \! 
            \xi_{\mu\mu_1\cdots \mu_n}(\bm{k})
           A_{\mu_1}(t) \cdots A_{\mu_{n}}(t).
           \label{eq:vnth}
\end{align}

The linear response $\delta J_{\mu} (t)$ can be nonzero in the EI state. However, because $\delta \bm{\phi}(t)=0$, the linear optical response does not reflect the dynamical effect of the order parameter.  
The second-order response $\delta^2 J_{\mu} (t)$ vanishes in the inversion symmetric system because the current has  odd parity under inversion. 
Hence, in order to see the dynamics of the excitonic order parameter, we need to evaluate the optical response at third order.  

Combining Eqs.~(\ref{eq:GF1st})-(\ref{eq:GF3rd}), (\ref{eq:Jnth}), and (\ref{eq:vnth}), we derive the third-order current $\delta^3 J_{\mu}$, which is comprised of the contributions diagrammatically described in Fig.~\ref{fig:diagJ}~\cite{parker2019}. 
Since the order parameter $\delta^2 \bm{\phi}(t)$ is included only in $\delta^2 h^A(\bm{k},t)$, the contributions 2b, 3b, and 3c are affected by the dynamical order parameter, which leads to the vertex correction terms (see Fig.~\ref{fig:diagJ}).  
Here, as an example, we derive the contribution from 3b but all  the THG susceptibilities are summarized in Appendix \ref{App:THGsus}. 
$v_{\mu}(\bm{k})$ and $G^0 \!*\! \delta\mathcal{H} \!*\! G^0  \!*\! \delta^2\mathcal{H} \!*\! G^0$ in $\delta^3 G$ gives $\delta^3 J_{\mu}(3\Omega)$ of 3b,
\begin{widetext}
\begin{align}
    \delta^3 J_{\mu} (3\Omega)_{3b}
        &= 3i  \left(\frac{e}{\hbar}\right) \sum_{\bm{k}} 
             \xi_{\mu}(\bm{k})
            \frac{1}{\hbar^2} \int \frac{d\omega}{2\pi}
            {\mathrm{tr}}\left[
                \sigma_z  G^{0}(\bm{k},\omega+3\Omega)  \delta h^A(\bm{k},\Omega)G^{0} (\bm{k},\omega+2\Omega) \delta^2 h^A(\bm{k},2\Omega)G^{0} (\bm{k},\omega)
            \right]^{<}.
\end{align}
Dividing as $\delta^3 J_{\mu} (3\Omega) = \delta^3 J^{0}_{\mu} (3\Omega) + \delta^3 J^{\rm vc}_{\mu} (3\Omega)$, the bare (0) and vertex correction (vc) terms are given by 
\begin{align}
    \delta^3 J^{0}_{\mu} (3\Omega)_{3b}
        &= 3i  \left(\frac{e}{\hbar}\right)^4
           \sum_{\mu_1,\mu_2,\mu_3}
              \sum_{\bm{k}}    \frac{1}{\hbar^2} \int \frac{d\omega}{2\pi}
            {\mathrm{tr}}\left[
               \sigma_z  G^{0}(\bm{k},\omega+3\Omega)  \sigma_zG^{0} (\bm{k},\omega+2\Omega) \sigma_z G^{0} (\bm{k},\omega)
            \right]^{<}
             \nonumber \\*
         &\hspace{170pt} \times
                     \xi_{\mu}(\bm{k})\xi_{\mu_1}(\bm{k})\xi_{\mu_2\mu_3}(\bm{k})
             A_{\mu_1}(\Omega) A_{\mu_2}(\Omega)A_{\mu_3}(\Omega),
\\
    \delta^3 J^{\rm vc}_{\mu} (3\Omega)_{3b}
        &= -3i U \left(\frac{e}{\hbar}\right)^4
           \sum_{\mu_1,\mu_2,\mu_3}\sum_a
              \sum_{\bm{k}}    \frac{1}{\hbar^2} \int \frac{d\omega}{2\pi}
            {\mathrm{tr}}\left[
                \sigma_z  G^{0}(\bm{k},\omega+3\Omega)  \sigma_zG^{0} (\bm{k},\omega+2\Omega) \sigma_a G^{0} (\bm{k},\omega)
            \right]^{<}
             \nonumber \\*
         &\hspace{170pt}\times
                    \xi_{\mu}(\bm{k})\xi_{\mu_1}(\bm{k})\Gamma_{\mu_2\mu_3}^a(2\Omega;\Omega,\Omega)
             A_{\mu_1}(\Omega) A_{\mu_2}(\Omega)A_{\mu_3}(\Omega),
\end{align}
respectively, where the vertex correction term $\delta^3 J^{\rm vc}_{\mu} (3\Omega)$ arises from the order parameter $\delta^2 \phi(2\Omega)$ in Eq.~(\ref{eq:OP2ndGamma}). 
The THG susceptibility may be defined as 
\begin{align}
&J^{(3)}_{\mu}(3\Omega)
    =  \delta^3 J_{\mu} (3\Omega)/3!
    = L^d \sum_{\mu_1,\mu_2,\mu_3}
    \chi^{(3)}_{\mu;\mu_1\mu_2\mu_3}(3\Omega; \Omega)
    A_{\mu_1}(\Omega)A_{\mu_2}(\Omega)A_{\mu_3}(\Omega), 
\end{align}
where $\chi^{(3)}_{\mu;\mu_1\mu_2\mu_3}(3\Omega; \Omega)=\chi^{(3)}_{\mu;\mu_1\mu_2\mu_3}(3\Omega; \Omega,\Omega,\Omega)$ and $L^d$ is the volume. 
Dividing $ \chi^{(3)}_{\mu;\mu_1\mu_2\mu_3}(3\Omega;\Omega)$ into  $\chi^{0;(3)}_{\mu;\mu_1\mu_2\mu_3}(3\Omega;\Omega)$ and $\chi^{{\rm vc};(3)}_{\mu;\mu_1\mu_2\mu_3}(3\Omega;\Omega)$, the bare and vertex correction terms of 3b are given by 
\begingroup\allowdisplaybreaks[1]
\begin{align}
    \chi^{0;(3,3b)}_{\mu;\mu_1\mu_2\mu_3}(3\Omega;\Omega)
        &= \frac{1}{2} i \left(\frac{e}{\hbar}\right)^4
            \frac{1}{\hbar^2} \int \! \frac{d\bm{k}}{(2\pi)^d} \! \int \! \frac{d\omega}{2\pi}
            \mathrm{tr}\left[
                \sigma_z G^0(\bm{k},\omega\!+\!3\Omega)
                \sigma_z G^0(\bm{k},\omega\!+\!2\Omega)
                \sigma_z G^0(\bm{k},\omega)
            \right]^<
                        \nonumber\\* &\hspace{235pt}\times
             \xi_{\mu}(\bm{k})
            \xi_{\mu_1}(\bm{k})
            \xi_{\mu_2\mu_3}(\bm{k}), 
          \\
    \chi^{{\rm vc};(3,3b)}_{\mu;\mu_1\mu_2\mu_3}(3\Omega;\Omega)
        &= -\frac{U}{2}i  \left(\frac{e}{\hbar}\right)^4
              \frac{1}{\hbar^2} \int \! \frac{d\bm{k}}{(2\pi)^d} \! \int \! \frac{d\omega}{2\pi}\sum_a
            {\mathrm{tr}}\left[
                \sigma_z  G^{0}(\bm{k},\omega\!+\!3\Omega)  \sigma_zG^{0} (\bm{k},\omega\!+\!2\Omega) \sigma_a G^{0} (\bm{k},\omega)
            \right]^{<}
            \nonumber\\* &\hspace{235pt}\times
                         \xi_{\mu}(\bm{k})\xi_{\mu_1}(\bm{k})\Gamma_{\mu_2\mu_3}^a(2\Omega;\Omega,\Omega), 
\end{align}
\endgroup
\end{widetext}
respectively. 
In the same way, we can derive the other THG susceptibilities and their formulas are summarized in Appendix \ref{App:THGsus}. 
Because of the vertex correction $\chi^{{\rm vc};(3)}_{\mu;\mu_1\mu_2\mu_3}(3\Omega;\Omega)$, the THG susceptibility can reflect the collective dynamics in the EI.


\begin{figure}[t]
    \begin{minipage}[t]{1.0\columnwidth}
        \begin{center}
        \includegraphics[bb=0 5 460.8 345.6,width=1\columnwidth]{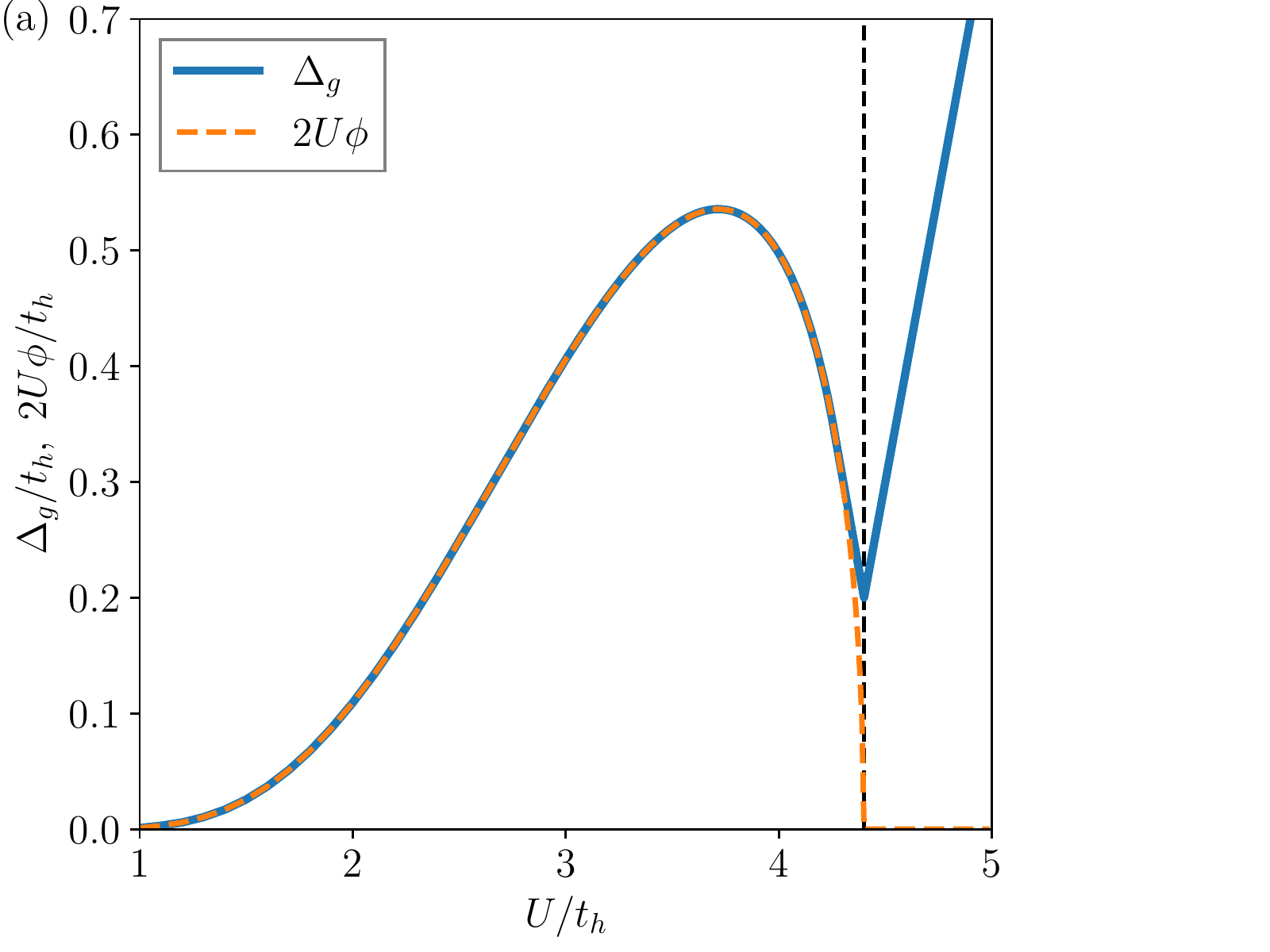}
        \end{center}
    \end{minipage}\\
    \begin{minipage}[t]{1.0\columnwidth}
        \begin{center}
        \includegraphics[bb=0 20 460.8 345.6,width=1\columnwidth]{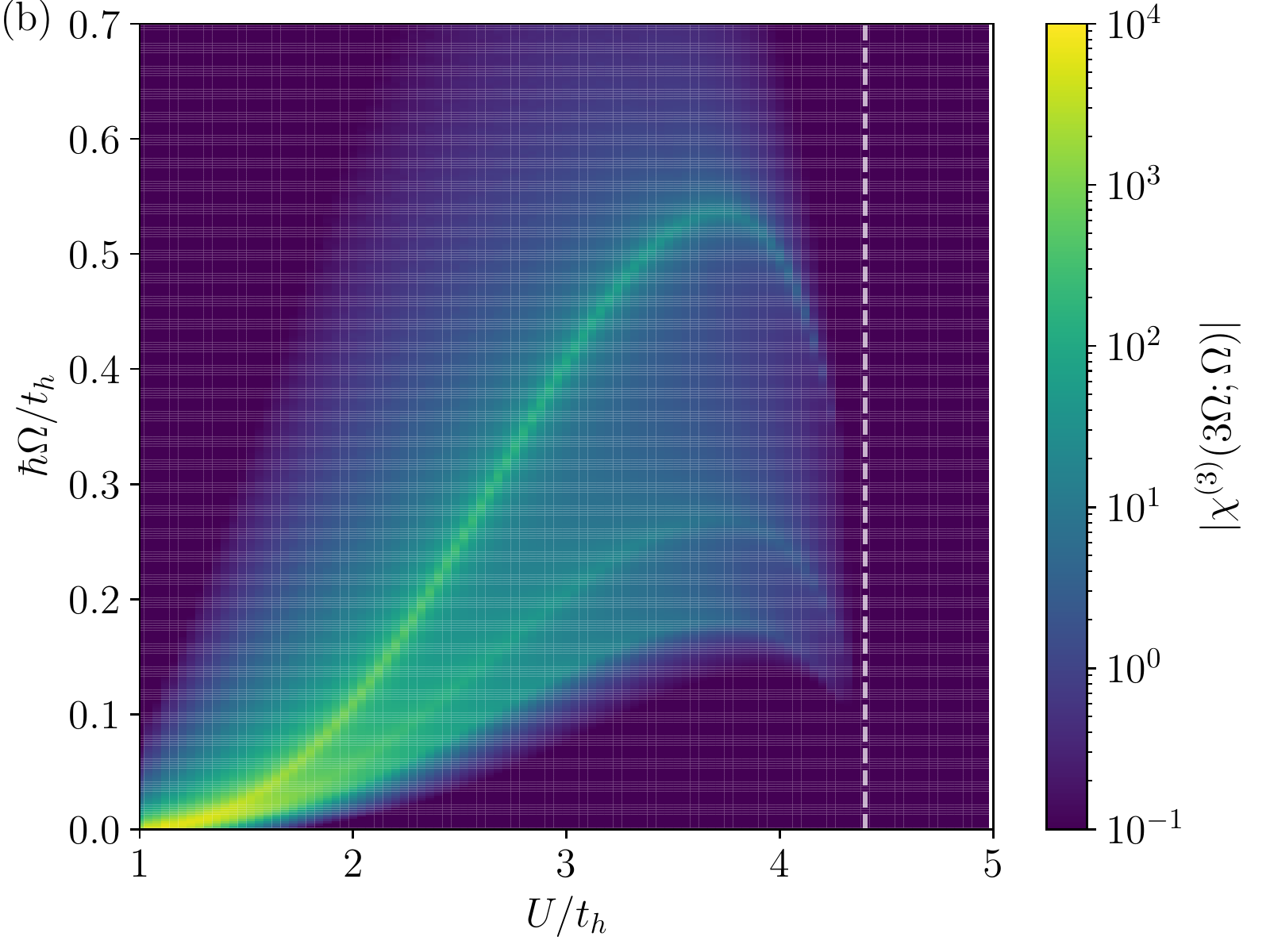}
        \end{center}
    \end{minipage}%
    \caption{(a) $U$ dependence of the excitonic order parameter $U \phi$ and the band gap $\Delta_g$ in the ground state. 
    The energy level difference is $D = \Delta_0 - \Delta_1 = 3.8 t_h$ and the vertical dashed line indicates the boundary of the EI phase ($U_c$). 
        (b)~THG susceptibility $|\chi^{(3)}_{x;xxx}(3\Omega;\Omega)|$ in the plane of $U$ and $\Omega$. $|\chi^{(3)}_{x;xxx}(3\Omega;\Omega)|$ is plotted in units of $(e/\hbar)^4 a^2t_h$ and the damping factor $\eta= 0.005t_h$ is used.}
    \label{fig:suscep_U_omega}
\end{figure}

\section{Third-harmonic generation} \label{sec:THG}
\subsection{THG susceptibility in the EI}

First, we show the THG susceptibility $\chi_{\mu;\mu_1\mu_2\mu_3}^{(3)}(3\Omega;\Omega)$ at zero temperature. 
Here we assume that the order parameter $\phi$ is real in the ground state without loss of generality and the external field is polarized along the $x$ direction. 
The polarization direction of the incident light does not change the main features of the THG susceptibility in the EI and the polarization dependence is discussed in  Appendix~\ref{App:pol}.
Here we set $t_h$ as a unit of the energy and plot the THG susceptibility $\chi_{x;xxx}^{(3)}(3\Omega;\Omega)$ in units of $(ea/\hbar)^4 t_h/a^d$ 
on the 2D square lattice ($d=2$). 

In order to see the change of the THG susceptibility from the BCS (small-$U$, semimetallic) regime to the BEC (large-$U$, semiconducting) regime, we plot the data by changing the Coulomb interaction $U$. 
Figure \ref{fig:suscep_U_omega}(a) shows the $U$ dependence of the band gap $\Delta_g$ and order parameter $\phi$ in the ground state. 
 While $\Delta_g = 2U\phi$ in the BCS regime, $\Delta_g > 2U \phi$ in the BEC semiconducting regime~\cite{seki2011}. 
 The order parameter vanishes above the phase boundary $U>U_c$, where the band gap $\Delta_g$ is larger than the exciton binding energy $E_B$~\cite{murakami2020}. 
Figure \ref{fig:suscep_U_omega}(b) is one of our main results, where we plot the magnitude of the THG susceptibility $|\chi_{x;xxx}^{(3)}(3\Omega;\Omega)|$ as a function of $U$.  
$\chi_{x;xxx}^{(3)}(3\Omega;\Omega)$ exhibits three peaks in the EI phase and their positions correspond to $\hbar \Omega = \Delta_g/3$, $\Delta_g / 2$, and $\Delta_g$ from the bottom. 
The THG response is strong in the BCS regime but it becomes less prominent with approaching the phase boundary $U_c$.  

\begin{figure*}[t]
    \begin{minipage}[t]{0.66\columnwidth}
        \begin{center}
        \includegraphics[bb=0 4.5 460.8 345.6,width=1\columnwidth]{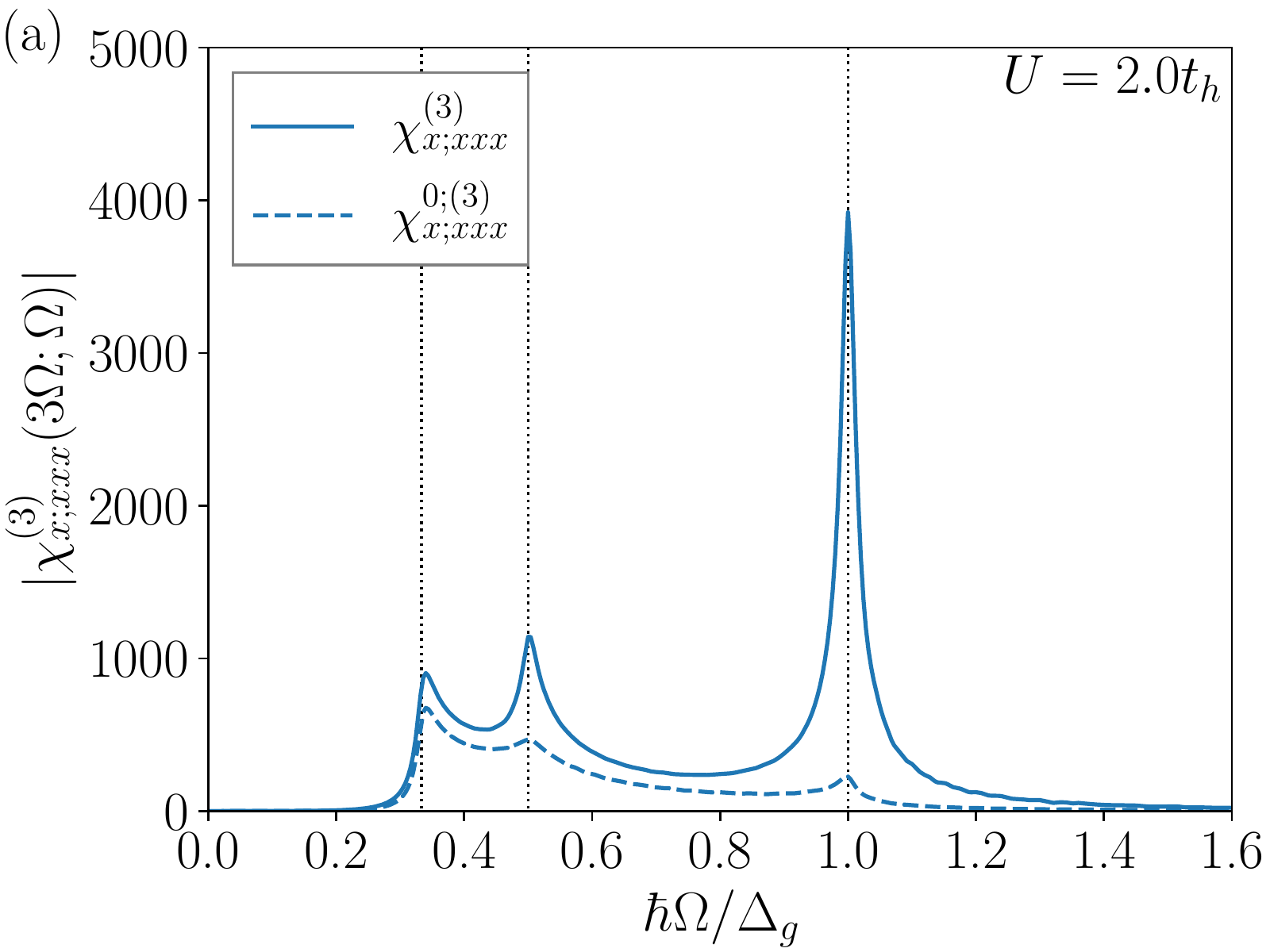}
        \end{center}
    \end{minipage}%
    \begin{minipage}[t]{0.66\columnwidth}
        \begin{center}
        \includegraphics[bb=0 4.5 460.8 345.6,width=1\columnwidth]{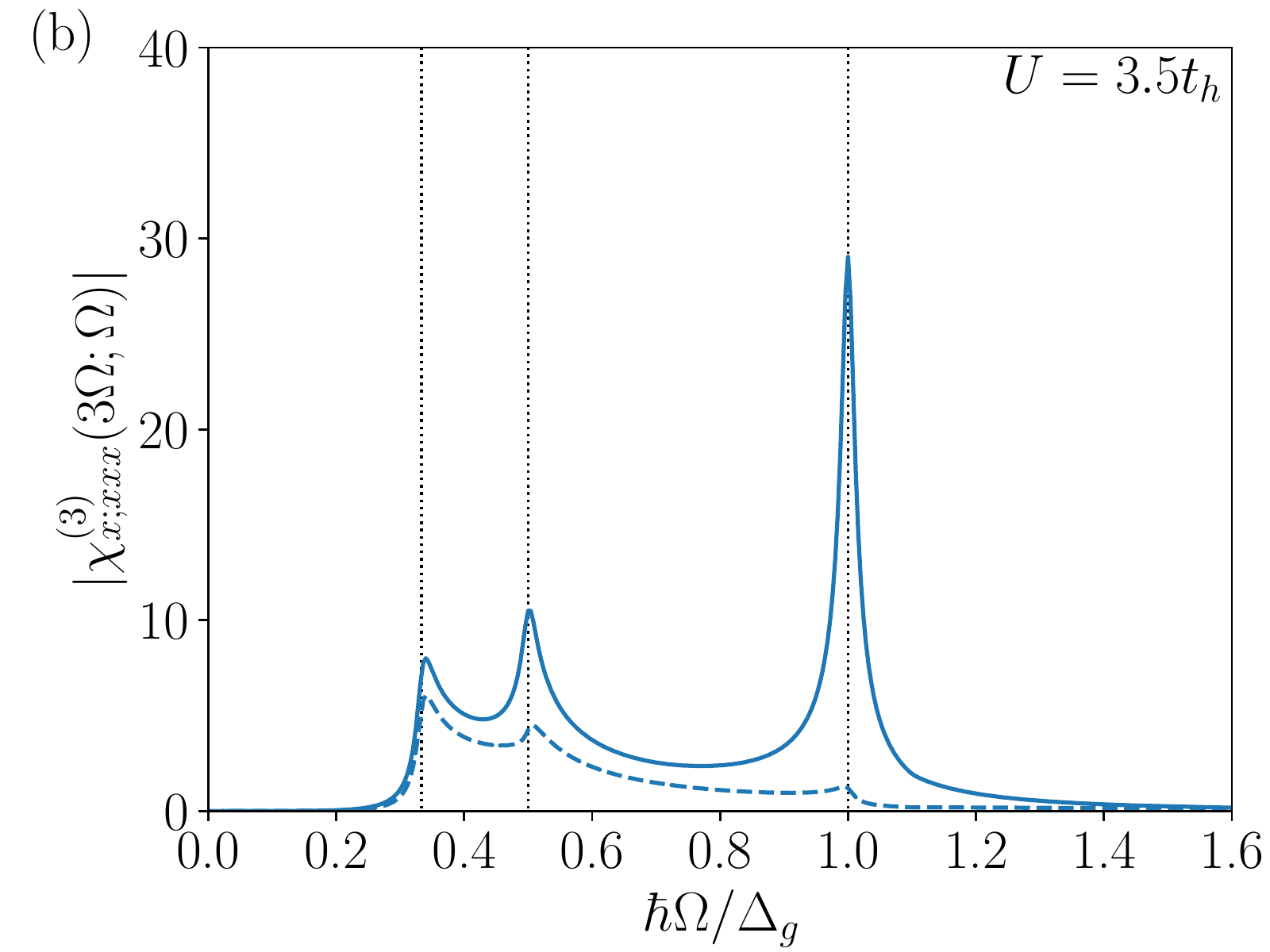}
        \end{center}
    \end{minipage}%
    \begin{minipage}[t]{0.66\columnwidth}
        \begin{center}
        \includegraphics[bb=0 4.5 460.8 345.6,width=1\columnwidth]{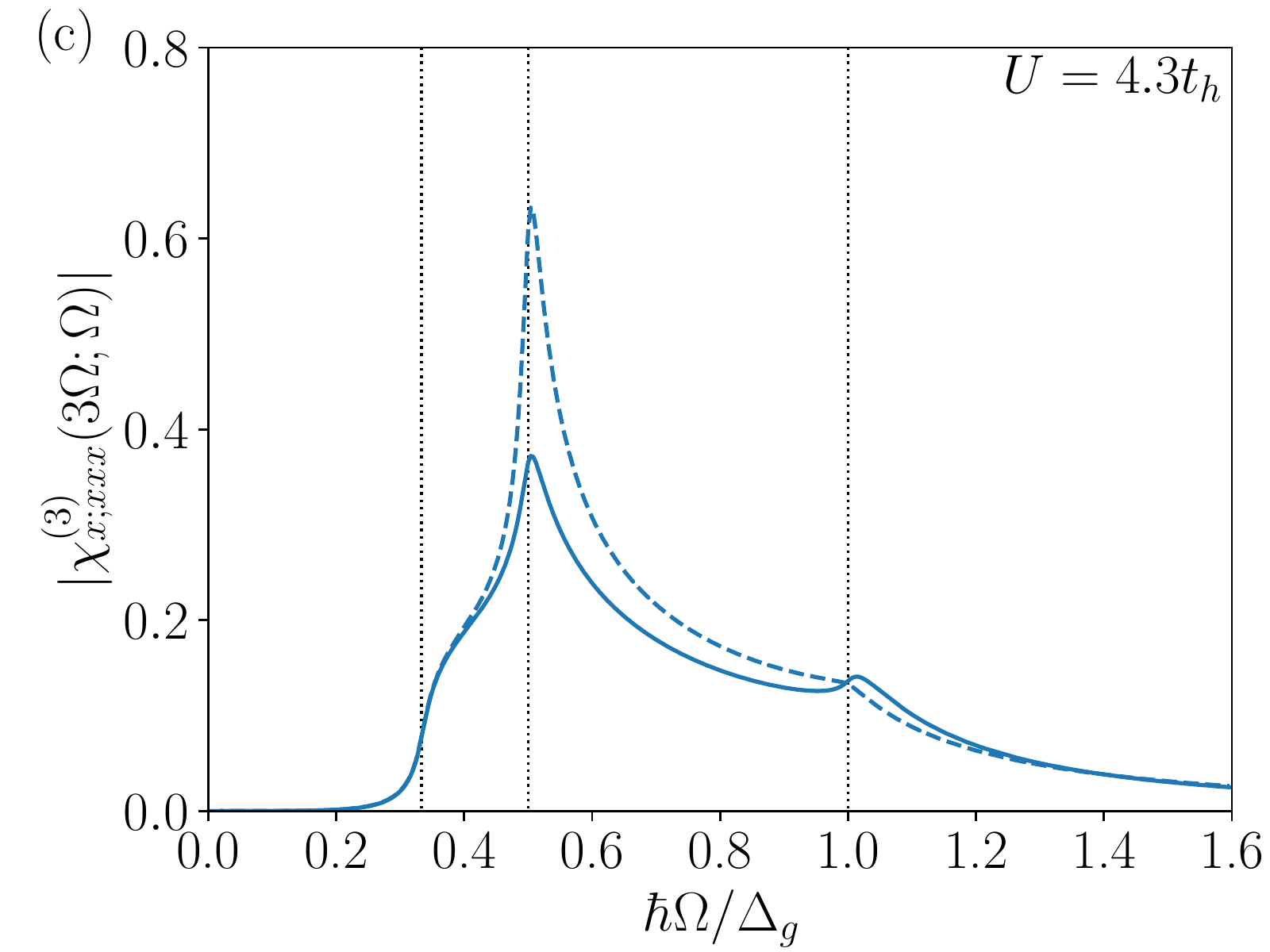}
        \end{center}
    \end{minipage}\\
    \begin{minipage}[t]{0.66\columnwidth}
        \begin{center}
        \includegraphics[bb=0 20 460.8 345.6,width=1\columnwidth]{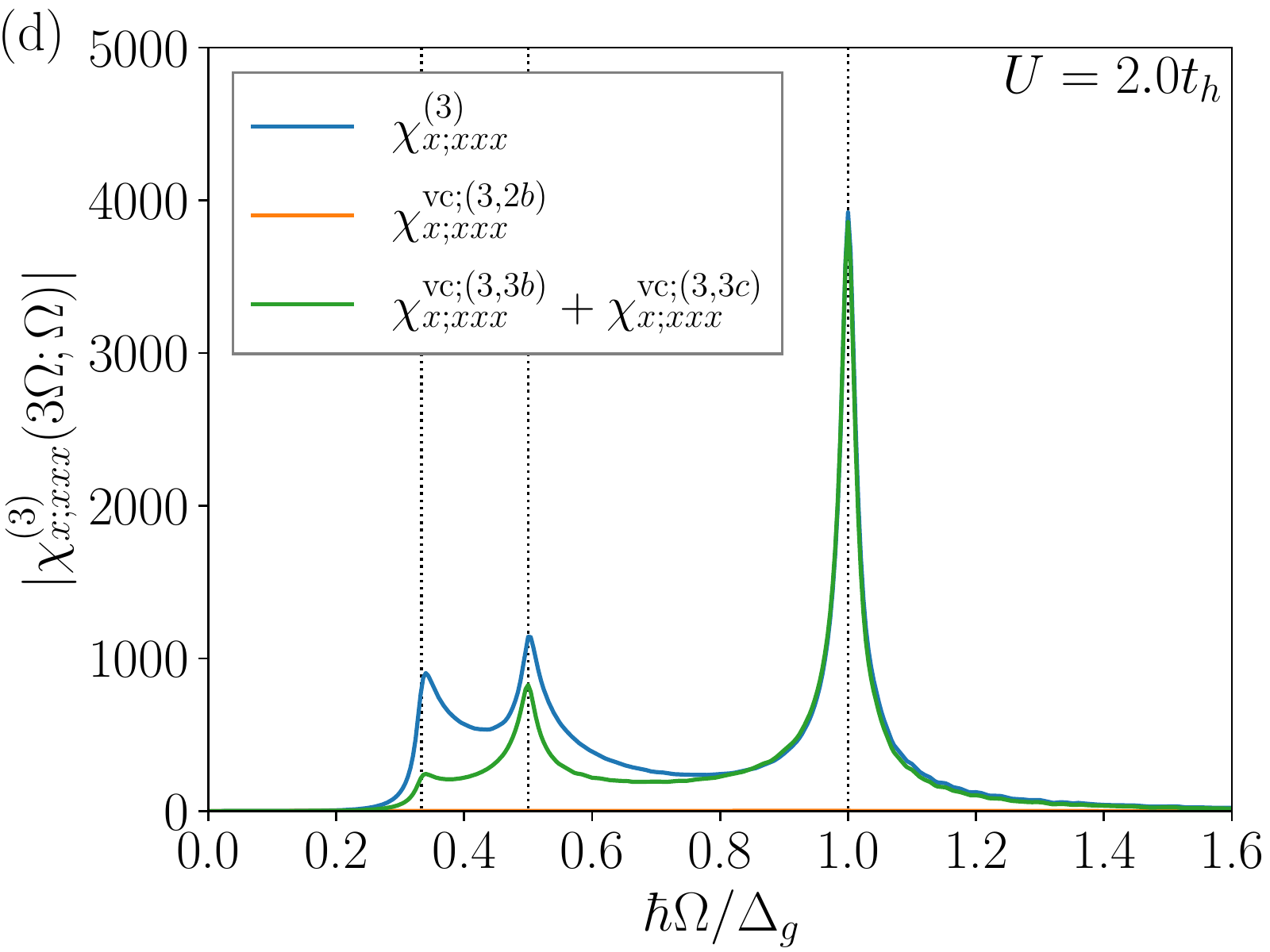}
        \end{center}
    \end{minipage}%
    \begin{minipage}[t]{0.66\columnwidth}
        \begin{center}
        \includegraphics[bb=0 20 460.8 345.6,width=1\columnwidth]{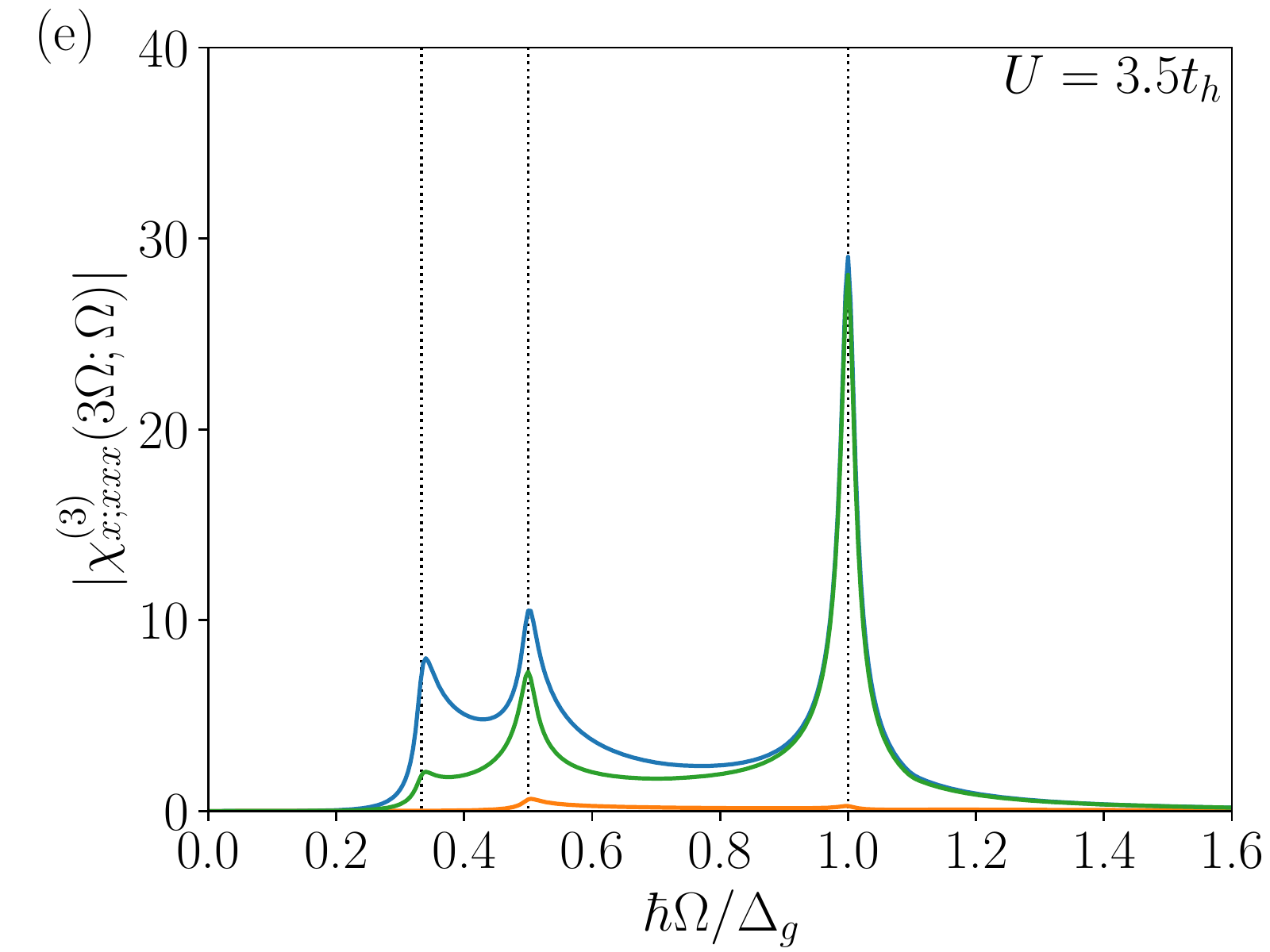}
        \end{center}
    \end{minipage}%
    \begin{minipage}[t]{0.66\columnwidth}
        \begin{center}
        \includegraphics[bb=0 20 460.8 345.6,width=1\columnwidth]{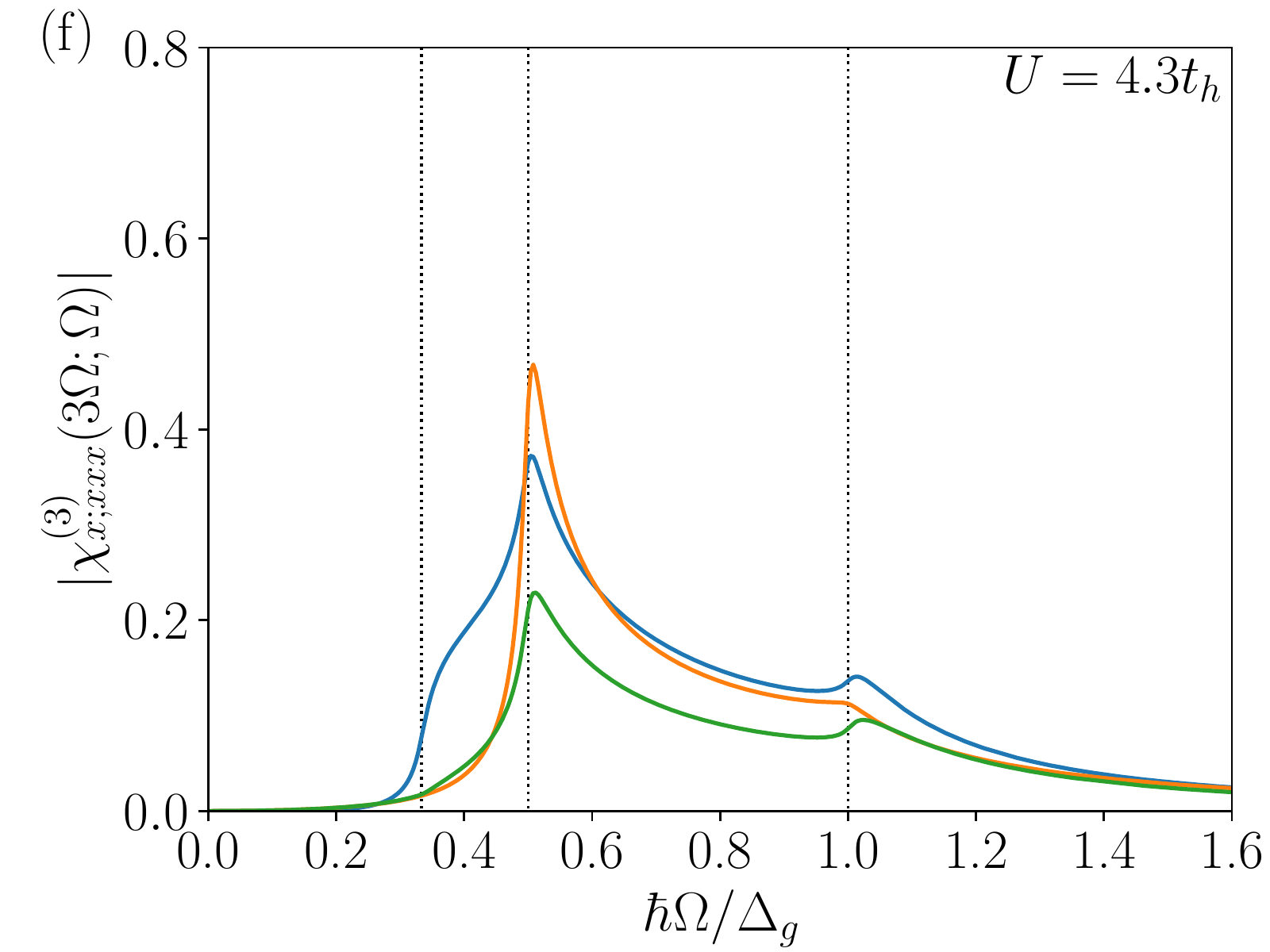}
        \end{center}
    \end{minipage}
    \caption{THG susceptibility $|\chi^{(3)}_{x;xxx}(3\Omega;\Omega)|$ at (a) and (d) $U=2t_h$ (BCS regime), (b) and (e) $U=3.5 t_h$ (intermediate regime), and (c) and (f) $U=4.3 t_h$ (BEC regime), where the horizontal axis is scaled by the band gap $\Delta_g$. 
   For comparison, the bare susceptibility $\chi^{0;(3)}_{x;xxx}(3\Omega;\Omega)$ is plotted in the upper panels [(a)-(c)] and the vertex corrections $\chi^{{\rm vc};(3,2b)}_{x;xxx}(3\Omega;\Omega)$ and $\chi^{{\rm vc};(3,3b)}_{x;xxx}(3\Omega;\Omega)+\chi^{{\rm vc};(3,3c)}_{x;xxx}(3\Omega;\Omega)$ (see Fig.~\ref{fig:diagJ}) are plotted in the lower panels [(d)-(f)]. 
                The vertical dotted lines indicates $\hbar \Omega = \Delta_g/3$, $\Delta_g/2$, and $\Delta_g$.  
             Here $\eta= 0.01\Delta_g$ is used while the other parameters are the same as in Fig.~\ref{fig:suscep_U_omega}.
             }
    \label{fig:suscep_omega}
\end{figure*}

\begin{figure}[t]
    \begin{center}
    \includegraphics[bb=0 40 460.8 345.6,width=\columnwidth]{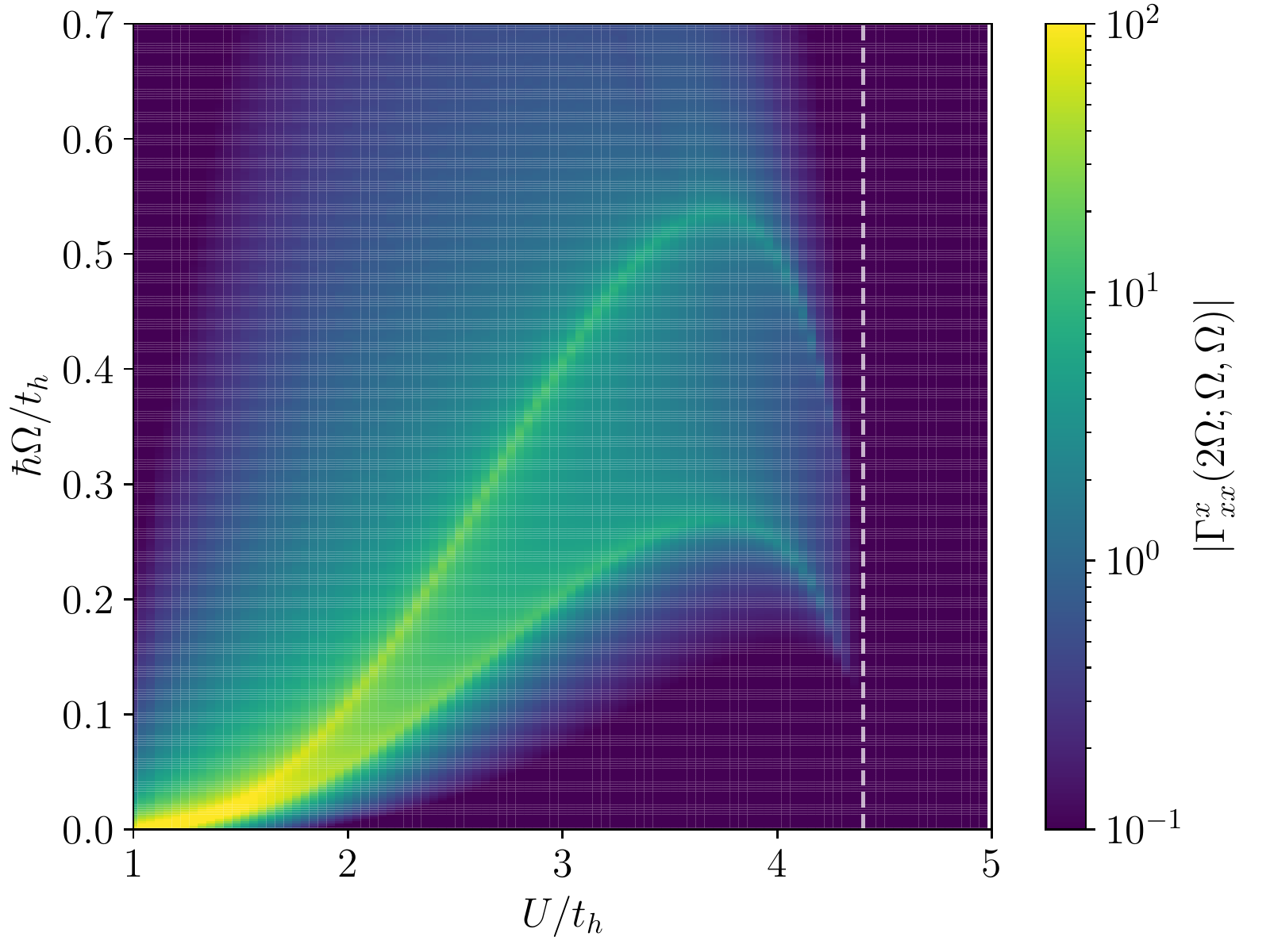}
    \end{center}
    \caption{Second-order response function for the order parameter $|\Gamma^x_{xx}(2\Omega;\Omega,\Omega)|$ in the plane of $U$ and $\Omega$. The vertical dashed line indicates the boundary of the EI phase ($U_c$). 
    Here $|\Gamma^x_{xx}(2\Omega;\Omega,\Omega)|$ is plotted in units of $a^2$ and the energy level difference $D = \Delta_0 - \Delta_1 = 3.8 t_h$ and the damping factor $\eta= 0.005t_h$ are used.}
    \label{fig:gamma_omega}
\end{figure}

\begin{figure}[t]
    \begin{center}
    \includegraphics[bb=0 40 460.8 345.6,width=0.95\columnwidth]{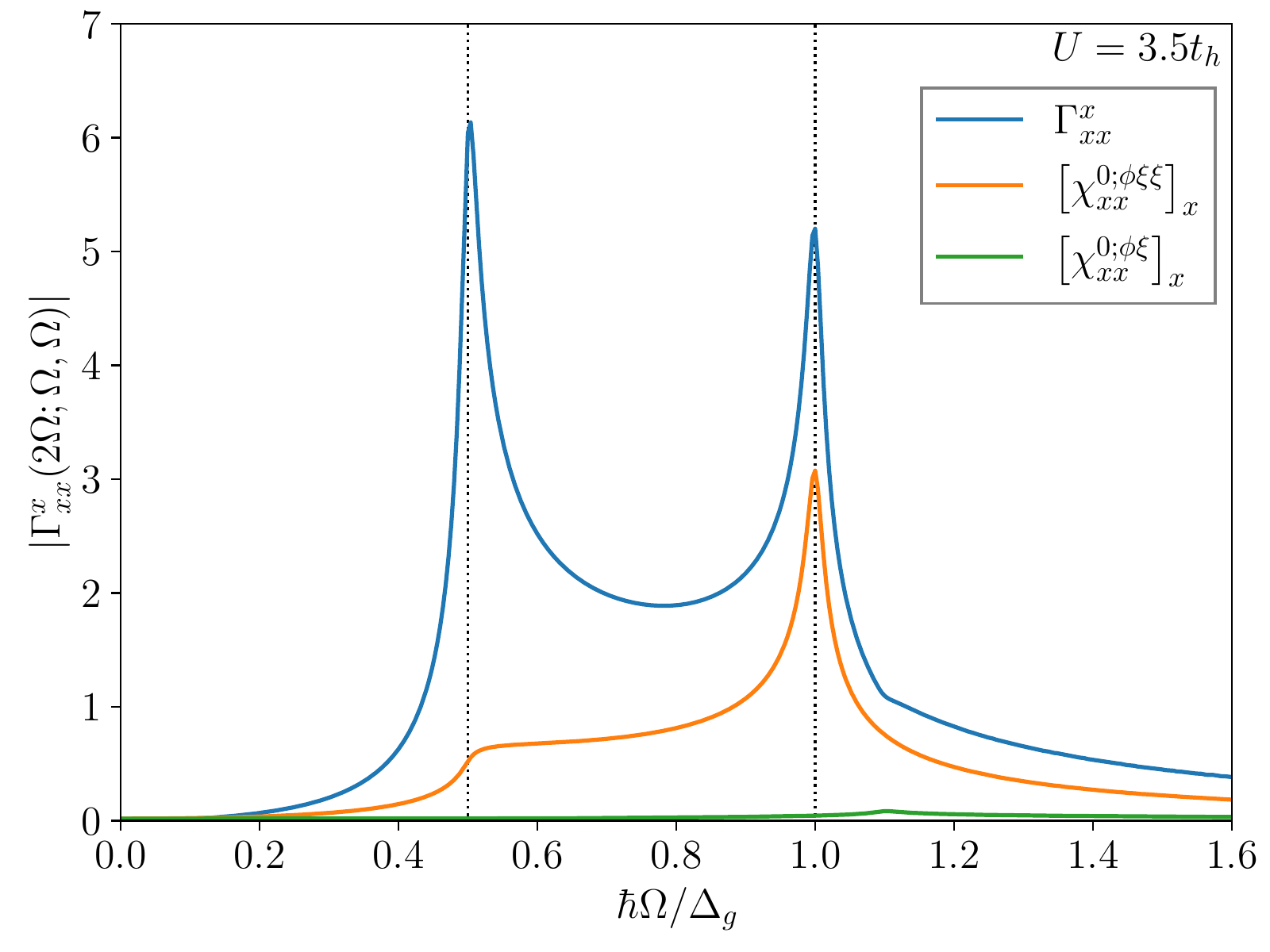}
    \end{center}
    \caption{Second-order response function for the order parameter $|\Gamma^x_{xx}(2\Omega;\Omega,\Omega)|$ at $U =3.5 t_h$, where the horizontal axis is scaled by the band gap $\Delta_g$. For comparison, the bare response functions $\left[\chi_{xx}^{0;\phi\xi\xi}(2\Omega;\Omega,\Omega)\right]_{x}$ and $\left[\chi_{xx}^{0;\phi\xi}(2\Omega;2\Omega)\right]_{x}$ are plotted. The vertical dotted lines indicates $\hbar \Omega = \Delta_g/2$, and $\Delta_g$. 
Here $\eta= 0.01\Delta_g$ is used while the other parameters are the same as in Fig.~\ref{fig:gamma_omega}.
}
    \label{fig:gamma_omega_int}
\end{figure}

Figure~\ref{fig:suscep_omega} shows the THG susceptibility as a function of $\hbar \Omega/\Delta_g$ in the BCS and BEC regimes. 
Here, in order to identify the contributions from the vertex correction, we plot the bare susceptibility $\chi^{0;(3)}_{x;xxx}(3\Omega;\Omega)$ in Figs.~\ref{fig:suscep_omega}(a)-\ref{fig:suscep_omega}(c) and the vertex corrections $\chi^{{\rm vc};(3,2b)}_{x;xxx}(3\Omega;\Omega)$ and $\chi^{{\rm vc};(3,3b)}_{x;xxx}(3\Omega;\Omega)+\chi^{{\rm vc};(3,3c)}_{x;xxx}(3\Omega;\Omega)$ in Figs.~\ref{fig:suscep_omega}(d)-\ref{fig:suscep_omega}(f).
All components of the bare THG susceptibilities and vertex corrections are presented in Appendix~\ref{App:THGsus}.  
As shown in Fig.~\ref{fig:suscep_omega}, while the THG susceptibility at $\hbar \Omega = \Delta_g / 3$ is mainly composed of the bare part $\chi^{0;(3)}_{x;xxx}(3\Omega;\Omega)$, the magnitude at $\hbar \Omega = \Delta_g / 2$ and $\Delta_g$ are modified by the vertex part $\chi^{{\rm vc};(3)}_{x;xxx}(3\Omega;\Omega)$. 
This indicates that, while THG at $\hbar \Omega = \Delta_g / 3$ is simply caused by bare three-photon excitation of the independent particle across the band gap $\Delta_g$, 
the order-parameter motions strongly contribute to THG at $\hbar \Omega = \Delta_g / 2$ and $\Delta_g$. 

In the BCS regime, the vertex correction enhances the THG susceptibility at both $\hbar \Omega = \Delta_g / 2$ and $\Delta_g$. 
In particular,  the peak at $\hbar \Omega = \Delta_g$ is outstanding. 
As shown in Figs.~\ref{fig:suscep_omega}(d) and \ref{fig:suscep_omega}(e), while the vertex correction in 2b (see Fig.~\ref{fig:diagJ}) is much smaller than the bare susceptibility, the vertex corrections in 3b and 3c  (see Fig.~\ref{fig:diagJ}) dominantly  enhance THG at $\hbar \Omega = \Delta_g / 2$ and bring the significant peak at $\hbar \Omega = \Delta_g$.   
Hence, we can observe the strong THG response due to the order-parameter motion, which is emergent in $\chi^{{\rm vc};(3,3b)}_{x;xxx}(3\Omega;\Omega)$ and $\chi^{{\rm vc};(3,3c)}_{x;xxx}(3\Omega;\Omega)$.  

In the BEC semiconducting regime, on the other hand, the peaks at $\hbar \Omega = \Delta_g / 2$ and $\Delta_g$ in the THG susceptibility become less prominent. 
As shown in Fig.~\ref{fig:suscep_omega}(c), $|\chi^{(3)}_{x;xxx}(3\Omega;\Omega)|$ at $\hbar \Omega = \Delta_g / 2$ is suppressed from the value of the bare susceptibility $|\chi^{0; (3)}_{x;xxx}(3\Omega;\Omega)|$.
The vertex correction in 3b and 3c is much smaller than its value in the BCS regime and is comparable to the vertex correction in 2b [see Fig.~\ref{fig:suscep_omega}(f)]. 
Since the vertex corrections are weak in the BEC regime, the resulting THG susceptibility does not show the collective excitonic nature strongly.   

Since the collective order parameter dynamics is important for THG in the EI, we show $\Gamma^{a}_{\mu_1\mu_2}(2\Omega;\Omega,\Omega)$ in Figs.~\ref{fig:gamma_omega} and \ref{fig:gamma_omega_int}, which is the response function of the MF parameter at second order in $\bm{A}(\Omega)$ [see Eq.~(\ref{eq:OP2ndGamma})]. 
Because we assume the order parameter $\phi$ is real in the ground state, the $a=x$ and $y$ components indicate the amplitude and phase oscillations of the order parameter, respectively. 
The $a=z$ component $\Gamma^{z}_{\mu_1\mu_2}(2\Omega;\Omega,\Omega)$ is zero when the hopping parameters satisfy $t_0+t_1=0$ at half-filling.  
While the $a=y$ (phase) component is nonzero, the $a=x$ (amplitude) component is dominant in particular in the BCS regime and here we present $\Gamma^{x}_{xx}(2\Omega;\Omega,\Omega)$. 
Figure~\ref{fig:gamma_omega} shows $\Gamma^{x}_{xx}(2\Omega;\Omega,\Omega)$ in the plane of $U$ and $\Omega$, where we find two peaks at $\hbar \Omega = \Delta_g/2$ and $\Delta_g$. 
The response of the amplitude oscillation is strong in the BCS regime but it becomes weaker with approaching the phase boundary $U_c$. 

In order to identify the origin of the two-peak structure, we compare $\Gamma^{x}_{xx}(2\Omega;\Omega,\Omega)$ with the bare response functions $\left[\chi_{xx}^{0;\phi\xi\xi}(2\Omega;\Omega,\Omega)\right]_{x}$ and $\left[\chi_{xx}^{0;\phi\xi}(2\Omega;2\Omega)\right]_{x}$ in Eqs.~(\ref{eq:chi_pxx}) and (\ref{eq:chi_px}), respectively. 
As shown in Fig.~\ref{fig:gamma_omega_int}, while the contribution from $\left[\chi_{xx}^{0;\phi\xi}(2\Omega;2\Omega)\right]_{x}$ is minor, $\left[\chi_{xx}^{0;\phi\xi\xi}(2\Omega;\Omega,\Omega)\right]_{x}$ exhibits the sharp peak at $\hbar \Omega =\Delta_g$, which is enhanced by the many-body correction in $\Gamma^{x}_{xx}(2\Omega;\Omega,\Omega)$. 
The response at $\hbar \Omega =\Delta_g/2$ is not prominent in the bare function,
but  the many-body correction $\left[1+U\tilde{\chi}^{\phi\phi}\right]^{-1}$  in $\Gamma^{x}_{xx}(2\Omega;\Omega,\Omega)$ gives rise to the resonant peak at $\hbar \Omega =\Delta_g/2$. 
Therefore, the origins of the peaks at $\hbar \Omega = \Delta_g/2$ and $\Delta_g$ are different, where the response at $\hbar \Omega = \Delta_g/2$ is originated by the many-body correction in $\Gamma^{x}_{xx}(2\Omega;\Omega,\Omega)$ while  the response at $\hbar \Omega = \Delta_g$ is mainly caused by the bare photon absorption described by the loop triangle diagram in Fig.~\ref{fig:diagOP}. 
Using Eq.~(\ref{eq:threeGinApp}) for the the loop triangle diagram, we find that $\left[\chi_{xx}^{0;\phi\xi\xi}(2\Omega;\Omega,\Omega)\right]_{x}$ includes the contribution represented by $(2\hbar\Omega)^{-1} \sum_{\bm{k}}\lambda(\bm{k})/\left[\hbar \Omega - |\bm{B}(\bm{k})|\right]$, which arises when one of two photon absorptions is resonant. 
This contribution gives rise to the prominent peak at $\hbar \Omega = \Delta_g$ in Fig.~\ref{fig:gamma_omega_int}. 

Since $\Gamma^{a}_{\mu_1\mu_2}(2\Omega;\Omega,\Omega)$ gives the vertex corrections in the THG susceptibility, two peaks observed in $\Gamma^{x}_{xx}(2\Omega;\Omega,\Omega)$ bring the resonant enhancement of THG at $\hbar \Omega = \Delta_g/2$ and $\Delta_g$.  
In the BEC regime, the vertex correction is small as shown in Fig.~\ref{fig:gamma_omega} and the resulting THG susceptibility does not exhibit significant peaks in comparison with the BCS-type EI.   
This is because the order parameter in the BEC-type EI is deeply stabilized at the bottom of the energy and is hard to deviate from its equilibrium value at second order in $\bm{A}(\Omega)$.  


\begin{figure}[t]
    \begin{minipage}[t]{\columnwidth}
        \begin{center}
        \includegraphics[bb=0 5 460.8 345.6,width=1\columnwidth]{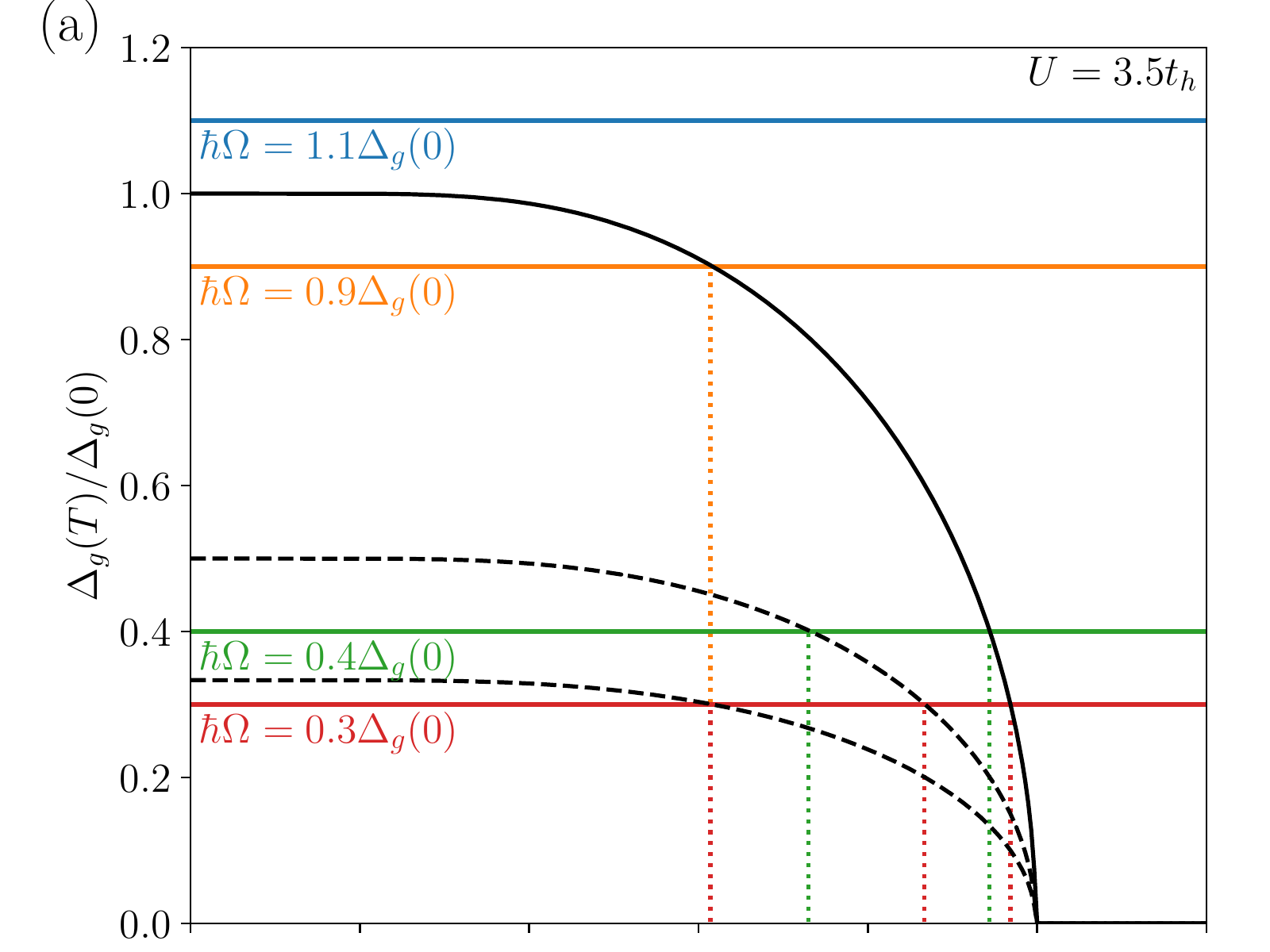}
        \end{center}
    \end{minipage}\\
    \begin{minipage}[t]{\columnwidth}
        \begin{center}
        \includegraphics[bb=0 20 460.8 518.4,width=1\columnwidth]{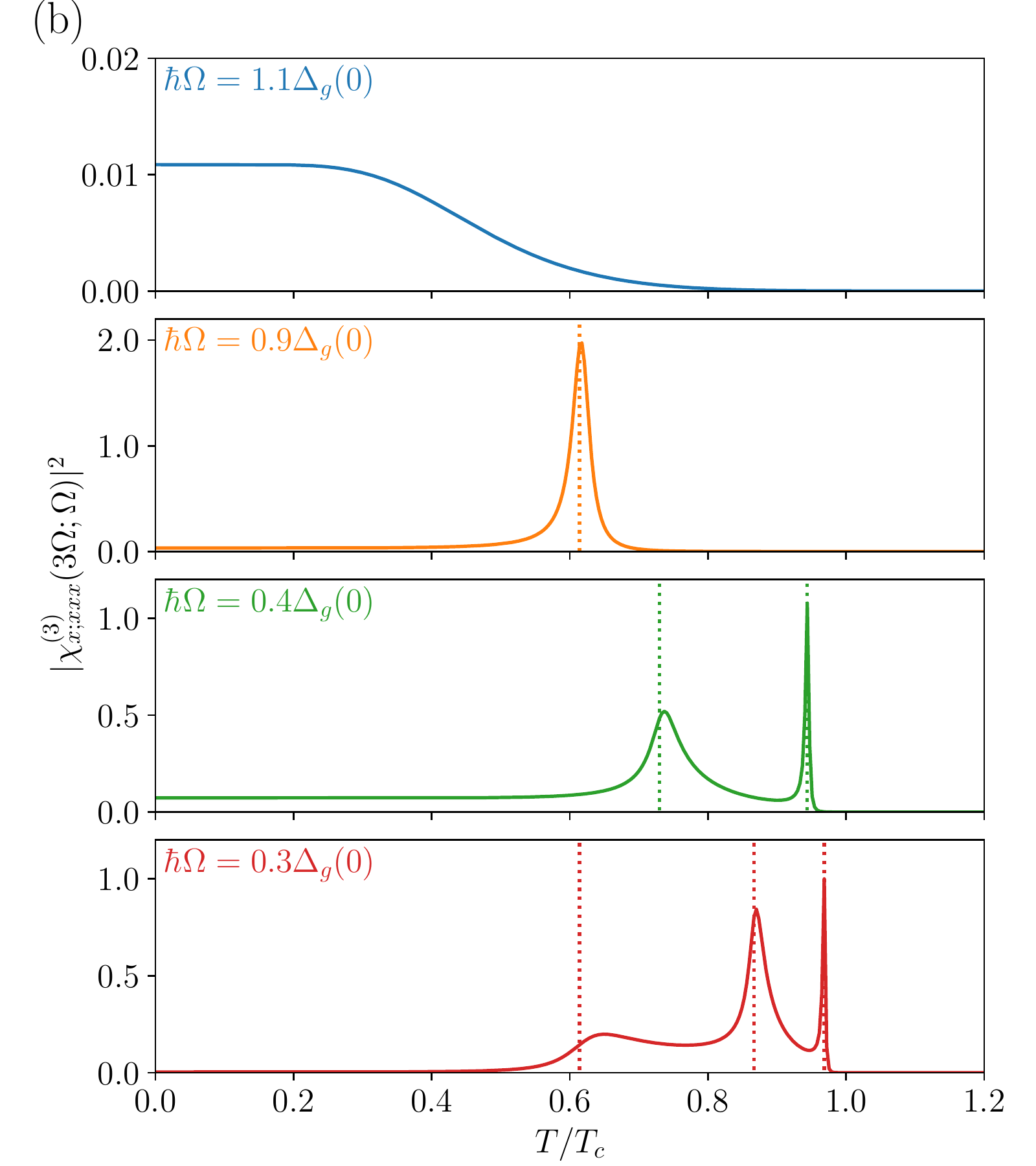}
        \end{center}
    \end{minipage}
    \caption{
   (a) Temperature dependence of the band gap $\Delta_g(T)$ at $U=3.5t_h$, where the dashed lines indicate $\Delta_g(T)/2$ and $\Delta_g(T)/3$. 
   The vertical dotted line indicates the temperature at which the order parameter and $\hbar \Omega$ cross.  
   (b)~Temperature dependence of $|\chi^{(3)}_{x;xxx}(3\Omega;\Omega)|^2$.   
   All intensities are normalized to the maximum value at $\hbar\Omega=0.3\Delta_g(T=0)$.
      The energy level difference $D = \Delta_0 - \Delta_1 = 3.8 t_h$ and the damping factor $\eta= 0.01\Delta_g(0)$ are used.
         }
    \label{fig:suscep_T_BCS-BEC}
\end{figure}

\begin{figure}[t]
    \begin{minipage}[t]{\columnwidth}
        \begin{center}
        \includegraphics[bb=0 5 460.8 345.6,width=1\columnwidth]{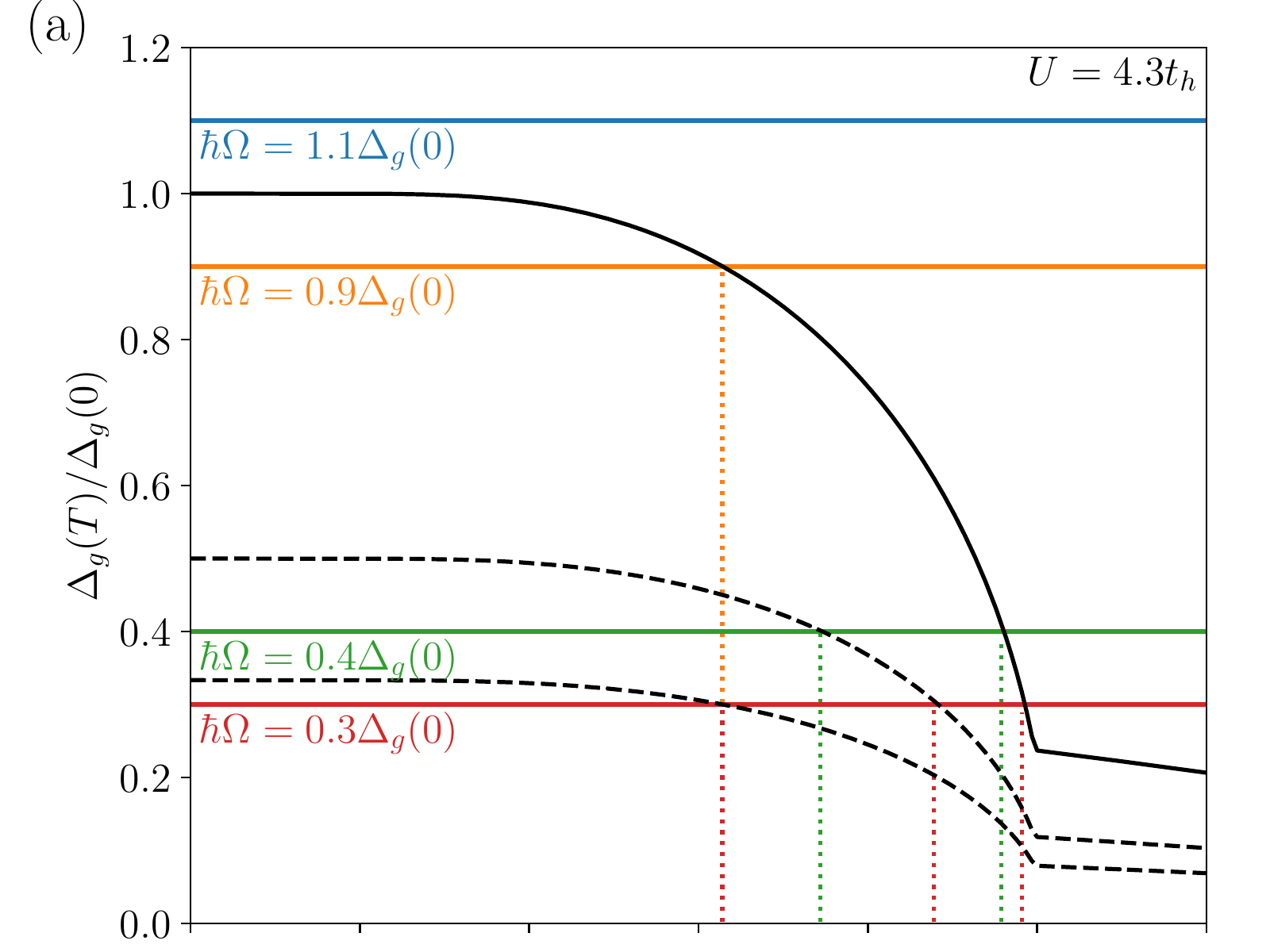}
        \end{center}
    \end{minipage}\\
    \begin{minipage}[t]{\columnwidth}
        \begin{center}
        \includegraphics[bb=0 20 460.8 518.4,width=1\columnwidth]{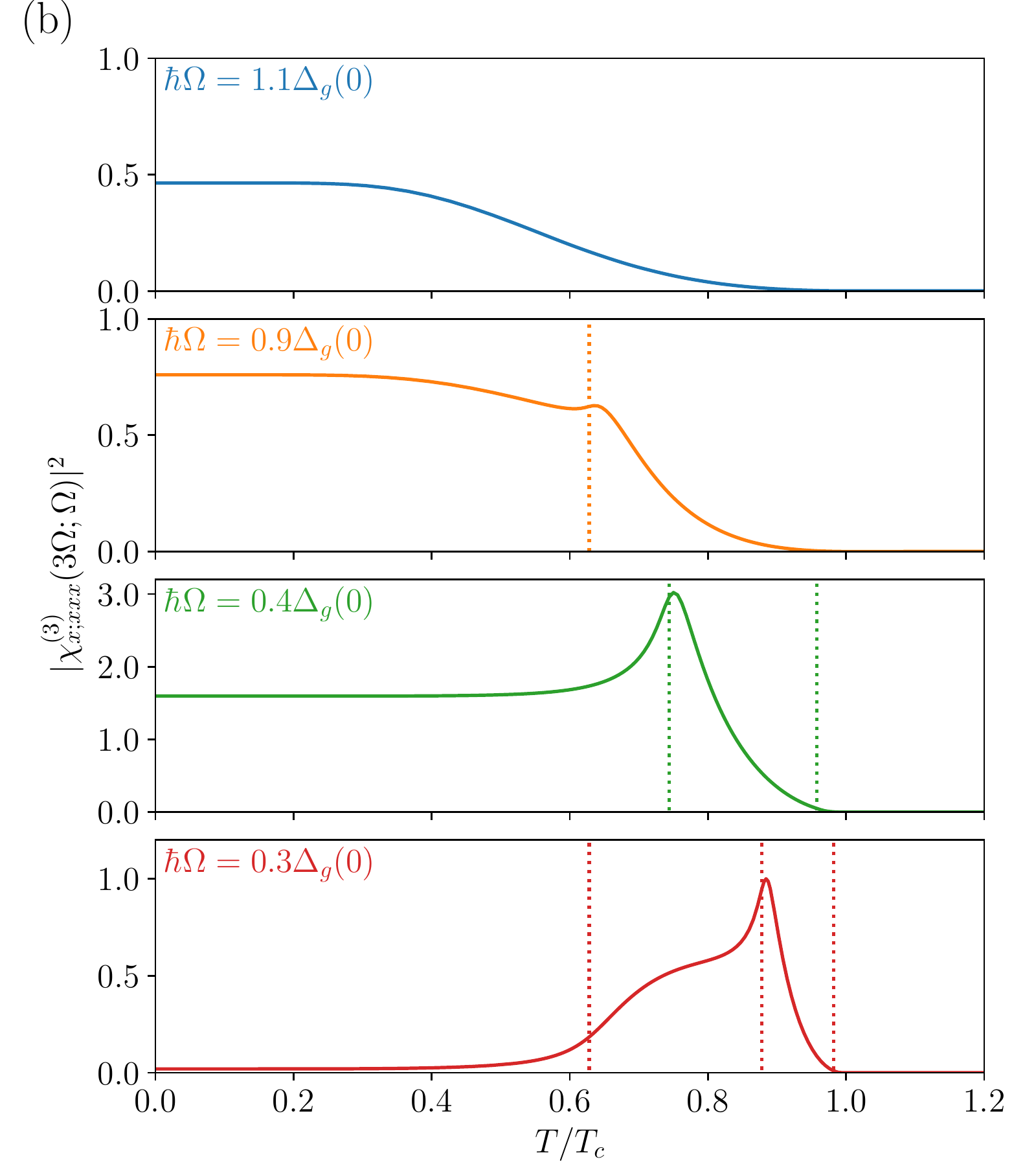}
        \end{center}
    \end{minipage}
    \caption{Same as Fig.~\ref{fig:suscep_T_BCS-BEC} but at $U=4.3t_h$ (BEC regime).}
    \label{fig:suscep_T_BEC}
\end{figure}

\subsection{Temperature dependence of the THG intensity}

In the  discussions of the Higgs-mode resonance in superconductors, the temperature profile of the THG intensity is  compared with experimental THG response~\cite{cea2016,schwarz2020,matsunaga2014,shimano2020}.  
Hence, we show the temperature dependence of the THG intensity for the EI~\cite{finiteT}. 
Here, we plot $|\chi^{(3)}_{x;xxx}(3\Omega;\Omega)|^2$ as the THG intensity $I_{\rm THG}$ since $I_{\rm THG} \propto |J^{(3)}_{\mu}(3\Omega)|^2$. 

Figure~\ref{fig:suscep_T_BCS-BEC} shows the results at $U=3.5 t_h$, where the strong THG responses are anticipated at $\hbar \Omega = \Delta_g$, $\Delta_g/2$, and $\Delta_g/3$ as shown in Fig.~\ref{fig:suscep_omega}(b).   
Actually, the temperature dependent THG intensity in Fig.~\ref{fig:suscep_T_BCS-BEC}(b) exhibits three peaks at the temperatures when $\hbar \Omega = 0.3 \Delta_g(T \!=\! 0)$ crosses $\Delta_g(T)$, $\Delta_g(T)/2$, and $\Delta_g(T)/3$, respectively.  
Associated with the number of the crossing points [see Fig.~\ref{fig:suscep_T_BCS-BEC}(a)], the THG intensities at $\hbar \Omega = 0.4 \Delta_g(0)$ and $0.9 \Delta_g(0)$ show two peaks and one peak, respectively, and the peak structure vanishes when $\hbar\Omega > \Delta_g(0)$. 
Therefore, when the BCS-like relation $\Delta_g = 2U\phi$ is well-satisfied [see Fig.~\ref{fig:suscep_U_omega}(a)], the number of the peaks in the temperature profile of the THG intensity decreases with increasing the light frequency $\Omega$.  

In the BEC semiconducting regime, on the other hand, the resonant peaks become less prominent as shown in Fig.~\ref{fig:suscep_omega}(c). 
Correspondingly, the THG intensity does not exhibit the strong resonant peak at the temperature when $\hbar \Omega$ crosses $\Delta_g(T)$ (see Fig.~\ref{fig:suscep_T_BEC}).  
While the THG intensity show the peak at $\hbar \Omega = \Delta_g(T)/2$ when $\hbar \Omega=0.3\Delta_g(0)$ and $0.4\Delta_g(0)$, 
the peak is not so sharp in comparison with THG in the BCS-type EI.


\section{Discussion and Summary}  \label{sec:summary}

While we studied THG in the purely electronic model, the low-temperature phases in the actual candidate materials, Ta$_2$NiSe$_5$ and TiSe$_2$, are associated with the structural phase transitions~\cite{disalvo1976,disalvo1986,holt2001,nakano2018}. 
Here we comment on effects of electron-phonon couplings briefly. 
The energy scale of lattice vibrations is usually much smaller than that of the band gap. 
Actually, the phonon frequency $\hbar \omega_{\rm ph} \sim 10-20$~meV while the band gap $\Delta_g \sim 200-300$~meV in Ta$_2$NiSe$_5$~\cite{larkin2017,larkin2018,kim2021,volkov2021npj,ye2021}.
In this condition, the phonon resonances appear at substantially low energies below the band gap, and the vertex corrections from phonons may be tiny (or negligible) in the region above the band gap because of the energy-scale  mismatch between the phonon frequency $\hbar \omega_{\rm ph}$ and the band gap $\Delta_g$~\cite{kaneko2021}. 
If the ordered state is purely phonon driven, the THG susceptibility is expected to be $\chi^{(3)}(3\Omega;\Omega) \sim \chi^{0; (3)}(3\Omega;\Omega)$ in the region above the band gap ($3\hbar \Omega > \Delta_g$) and we may not observe the resonant peaks shown in Figs.~\ref{fig:suscep_omega} and \ref{fig:suscep_T_BCS-BEC}. 
Therefore, the resonant peaks we find can be a smoking gun for the identification of the excitonic order.  
If the THG intensities in experiments exhibit the temperature profile as shown in Fig.~\ref{fig:suscep_T_BCS-BEC}, we may conclude that the ordered state is a BCS-type EI.  
However, if it is not observed, there may be two possibilities: (1) An ordered state is dominantly phonon driven as speculated here or (2) an ordered state is a BEC-type (strong-coupling) EI as shown  in Fig.~\ref{fig:suscep_T_BEC} since $\chi^{(3)}(3\Omega;\Omega) \sim \chi^{0; (3)}(3\Omega;\Omega)$ [see, e.g.,~Fig.~\ref{fig:suscep_omega}(c)]. 
If we can drive the collective motion more actively by a strong electric field, we might observe the nonlinear excitonic collective nature even in the BEC-type  EI and distinguish it from the phonon-driven case. 
In order to address the above issue, one needs to make detailed analyses and calculations of  high-harmonic generation in an electron-phonon coupled model or realistic models for the candidate materials, which will be important extensions of the present study in the future. 

To conclude, we have investigated THG in the EI state described in the two-band spinless model. 
We have derived the THG susceptibility taking into account the vertex corrections and have shown that the order-parameter motion is activated at second order of the external field and its effects arise in THG. 
We have found that the THG susceptibility exhibit three peaks at $\hbar \Omega = \Delta_g/3$, $\Delta_g/2$, and $\Delta_g$.  
While THG at $\Delta_g/3$ is simply caused by bare three-photon excitation of the independent particle across the band gap, the latter two peaks are attributed to the 
dynamical order parameter $\delta^2 \phi(2\Omega)$ activated at second order, where the resulting resonant peaks are prominent in the BCS regime but they become less prominent in the BEC  regime. 
We have identified that the motion of the order parameter at $\hbar\Omega = \Delta_g$ is mainly caused by the bare photon absorption while the mode at $\Delta_g/2$ is originated from the many-body correction.    
We have also demonstrated that  the resonant peak caused by the collective motion is observable in the temperature profile of the THG intensity. 
Our finding suggests that the THG measurement is promising for detecting the excitonic collective nature of materials.


\begin{acknowledgments}
The authors acknowledge K. Sugimoto and Y. Murakami for fruitful discussion. 
This work was supported by Grants-in-Aid for Scientific Research from JSPS, KAKENHI Grants
No. JP17K05530, JP18K13509, and No. JP20H01849.
T.K. was supported by the JSPS Overseas Research Fellowship. 
The diagrams in our figures
are produced using JaxoDraw~\cite{binosi2004}.
\end{acknowledgments}


\appendix 


\section{Green's function}\label{App:GF}

The nonequilibrium Green's function is defined as 
\begin{align}
    G(\bm{k},t,t') & = \left[\begin{array}{cc}
        G^T(\bm{k},t,t') & G^<(\bm{k},t,t') \\
        G^>(\bm{k},t,t') & G^{\tilde{T}}(\bm{k},t,t')
    \end{array}\right]. 
\end{align}
Here each component is $2\times 2$ matrix and 
\begin{align}
    \left[G^T(\bm{k},t,t')\right]_{\alpha\beta}
        &= -i\left\langle T\left[
                \hat{c}_{\bm{k},\alpha}(t)\hat{c}^{\dagger}_{\bm{k},\beta}(t')
            \right] \right\rangle ,\\
    \left[G^{\tilde{T}}(\bm{k},t,t')\right]_{\alpha\beta}
        &= -i\left\langle {\tilde{T}}\left[
                \hat{c}_{\bm{k},\alpha}(t)\hat{c}^{\dagger}_{\bm{k},\beta}(t')
            \right] \right\rangle , \\
    \left[G^<(\bm{k},t,t')\right]_{\alpha\beta}
        &= i\left\langle
                \hat{c}^{\dagger}_{\bm{k},\beta}(t')\hat{c}_{\bm{k},\alpha}(t)
            \right\rangle , \\
    \left[G^>(\bm{k},t,t')\right]_{\alpha\beta}
        &= -i\left\langle
                \hat{c}_{\bm{k},\alpha}(t)\hat{c}^{\dagger}_{\bm{k},\beta}(t')
            \right\rangle ,
\end{align}
where $T$ ($\tilde{T}$) indicates the time(anti-time)-ordered product. 

For a general nonequilibrium correlation function [e.g., $X(t,t')=G(\bm{k},t,t')$] defined as 
\begin{align}
    X(t,t')
    &= \left[\begin{array}{cc}
        X^T(t,t') & X^<(t,t') \\
        X^>(t,t') & X^{\tilde{T}}(t,t')
    \end{array}\right]
    \notag \\
    &= \left[\begin{array}{cc}
        X^{++}(t,t') & X^{+-}(t,t') \\
        X^{-+}(t,t') & X^{--}(t,t')
    \end{array}\right], 
\end{align}
the retarded/advanced component is given by 
\begin{align}
    X^{R/A}(t,t') &\equiv X^T(t,t')-X^{</>}(t,t'). 
\end{align}
The matrix multiplication between $X(t,t')$ and $Y(t,t')$ is defined by 
\begin{align}
    (X*Y)^{\zeta\zeta'}(t,t')
    \equiv \sum_{\zeta_1=\pm}\zeta_1 \int_{-\infty}^{\infty} dt_1
            X^{\zeta\zeta_1}(t,t_1)
            Y^{\zeta_1\zeta'}(t_1,t'), 
\end{align}
where $\zeta_1=-$ ($+$) arises from the contour $\mathcal{C}^-$: $t_1= \infty \rightarrow t_1=-\infty$ ($\mathcal{C}^+$: $t_1=-\infty \rightarrow t_1=\infty$). 
The lesser component of the product $X_1*X_2* \cdots *X_n$ follows the Langreth's rule 
\begin{align}
    &(X_1 * X_2 * \cdots * X_n)^<(t,t')
     \\
   & = \sum_{i=1}^n\int_{-\infty}^{\infty} dt_1 \cdots dt_{n-1}
            X_1^R(t,t_1)X_2^R(t_1,t_2)\cdots
            \notag \\
       & \qquad \qquad \qquad \qquad \quad \times 
            X_i^<(t_{i-1},t_i)\cdots
            X_n^A(t_{n-1},t'). 
            \notag 
\end{align}

The nonequilibrium Green's function satisfies 
\begin{align}
    (G*G^{-1})(\bm{k},t,t')
    = I(t,t')
    \equiv \left[\begin{array}{cc}
        \sigma_0& 0 \\
        0 & -\sigma_0
    \end{array}\right]\delta(t-t'), 
\end{align}
where we find 
\begin{align}
    [G^{-1}(\bm{k},t,t')]^{\zeta\zeta}
        = \zeta\left[  i\sigma_0\frac{\partial}{\partial t}\delta(t-t') - \frac{1}{\hbar}h^A(\bm{k},t)\delta(t-t')  \right]
\end{align}
and $[G^{-1}(\bm{k},t,t')]^{+-}= [G^{-1}(\bm{k},t,t')]^{-+}=0$. 
Then the deviation from equilibrium is given by $\delta^n \mathcal{H} = -\delta^nG^{-1}$. 
The variation of $G^{-1}*G = I$ with respect to $\bm{A}$ gives rise to the equations 
\begin{align}
    &{G^{0}}^{-1}*\delta{G} - \delta\mathcal{H}*G^0=0,  \notag \\
    &{G^{0}}^{-1}*\delta^2{G} - 2\delta\mathcal{H}*\delta{G} - \delta^2\mathcal{H}*G^0 = 0, \notag \\
    & \qquad \vdots \notag 
\end{align}
sequentially. By multiplying $G^{0}$ from left, we obtain the Green's functions 
\begin{align}
    \delta G
        &= G^0*\delta\mathcal{H}*G^0 ,\notag \\
    \delta^2 G
        &= 2 G^0 * \delta\mathcal{H} * G^0 * \delta\mathcal{H} * G^0
          + G^0 * \delta^2\mathcal{H} * G^0 ,\notag \\
        & \;\; \vdots \; . \notag
\end{align}
Combining the above Green's function and Langreth's rule, for example, the lesser component of $\delta{G}=G^0*\delta\mathcal{H}*G^0$  is given by 
\begin{align}
    &\delta{G}^<(\bm{k},t,t') \nonumber\\
    &= \frac{1}{\hbar}\int dt_1
            G^{0,R}(\bm{k},t-t_1)
            \delta h^A(\bm{k},t_1)
            G^{0,<}(\bm{k},t_1-t')
    \nonumber\\
    &+ \frac{1}{\hbar}\int dt_1
            G^{0,<}(\bm{k},t-t_1)
            \delta h^A(\bm{k},t_1)
            G^{0,A}(\bm{k},t_1-t'). 
\end{align}
In the same way we can derive the Green's function $\delta^n{G}^<$ at $t=t'$, which is used for the estimation of the time-dependent quantities $\delta^n \bm{\phi}(t)$ and $\delta^n J_{\mu}(t)$ [e.g., Eqs.~(\ref{eq:OP2nd}) and (\ref{eq:Jnth})]. 
For $\bm{A}(t) = \bm{A}(\Omega)e^{-i\Omega t} + {\rm c.c.}$, the Fourier coefficient of $\delta{G}^<(\bm{k},t,t)$  is given by 
\begin{align}
    &\delta{G}^<(\bm{k},\Omega) \nonumber\\
    &= \frac{1}{\hbar}\int \frac{d\omega}{2\pi}
            G^{0,R}(\bm{k},\omega+\Omega)
            \delta h^A(\bm{k},\Omega)
            G^{0,<}(\bm{k},\omega)
    \nonumber\\
    &+ \frac{1}{\hbar}\int \frac{d\omega}{2\pi}
            G^{0,<}(\bm{k},\omega+\Omega)
            \delta h^A(\bm{k},\Omega)
            G^{0,A}(\bm{k},\omega). 
\end{align}
Following the Langreth's rule, we summarize the terms in the right-hand side as 
\begin{align}
    \delta{G}^<(\bm{k},\Omega) 
    = \frac{1}{\hbar}\int \! \frac{d\omega}{2\pi}
            \left[
            G^{0}(\bm{k},\omega \!+\! \Omega)
            \delta h^A(\bm{k},\Omega)
            G^{0}(\bm{k},\omega)
            \right]^<.
\label{eq:GF1st_A}
\end{align}

\section{MF parameter at odd order} \label{MFodd}
Here we show vanishing of the order parameter at the odd order in the external field. 
For example, combining Eqs.~(\ref{eq:OPGF}) and (\ref{eq:GF1st_A}), the order parameter at the first order is given by
\begin{align}
    \delta \phi_{a}(\Omega)
            =
 -i\frac{1}{2N}\sum_{\bm{k}}   \frac{1}{\hbar} \int \frac{d\omega}{2\pi}
                         {\mathrm{tr}}\bigl[\sigma_a  G^{0}(\bm{k},\omega+\Omega) &
                         \notag \\ 
                         \times \delta h^A(\bm{k},\Omega)G^{0} (\bm{k},\omega) & \bigr]^{<}     , 
\label{eq:OP1stSCE}                   
\end{align}
where
$\delta h^A(\bm{k},\Omega) =  - U \delta \bm{\phi}(\Omega)  \cdot  \bm{\sigma} + (e/\hbar)  \sum_{\mu_1} \! \xi_{\mu_1}(\bm{k}) A_{\mu_1}(\Omega) \,\sigma_z $. 
However, because $G^{0} (-\bm{k},\omega) = G^{0} (\bm{k},\omega)$ and $\xi_{\mu}(-\bm{k}) = - \xi_{\mu}(\bm{k})$, the term originated from $\xi_{\mu}(\bm{k}) A_{\mu}(\Omega)$ is an odd function for $\bm{k}$ and vanishes due to the $\bm{k}$ summation in Eq.~(\ref{eq:OP1stSCE}). 
Then we find
\begin{align}
\delta \bm{\phi} (\Omega) = -U \tilde{\chi}^{0;\phi\phi}(\Omega)\delta  \bm{\phi} (\Omega), 
\end{align}
where $\tilde{\chi}^{0;\phi\phi}(\Omega)$ is the same function with Eq.~(\ref{eq:sus_phiphi}).
Since the solution of this equation is $\delta \bm{\phi} (\Omega)=\bm{0}$, the order parameter at the first order vanishes. 
In the same way, the order parameters at higher odd orders also vanish.


\section{$\omega$ integral} \label{App:integw}
Here we consider the $\omega$ integral in 
\begin{align}
    \int \frac{d\omega}{2\pi}
        {\mathrm{tr}}\left[
            \sigma_a  G^{0}(\bm{k},\omega+\Omega_1) \sigma_b G^{0} (\bm{k},\omega+\Omega_2) \cdots
        \right]^{<}.
\end{align}
Using the bare Green's functions
\begin{align}
    G^{0,<} (\bm{k},\omega)
        &= \sum_{\nu=\pm}  b_{\nu}(\bm{k})F_{\nu}^{<}(\bm{k},\omega), \\
    G^{0,R/A} (\bm{k},\omega)
        &= \sum_{\nu=\pm}  b_{\nu}(\bm{k}) F_{\nu}^{R/A}(\bm{k},\omega), 
\end{align}
where 
\begin{align}
    &F_{\nu}^{<}(\bm{k},\omega)
        \equiv
            2\pi i f(E_{\nu}(\bm{k})) 
            \delta\left(\omega-\omega_{\nu}(\bm{k})\right), \\
    &F_{\nu}^{R/A}(\bm{k},\omega)
        \equiv
            \frac{1}{\omega-\omega_{\nu}(\bm{k}) \pm i 0^+}, 
\end{align}
we can divide the integrand into the trace part ${\mathrm{tr}}[\sigma_a  b_{\nu_1}(\bm{k}) \sigma_b b_{\nu_2}(\bm{k}) \cdots]$ and the $\omega$-integral  part $\int \! \frac{d\omega}{2\pi}\left[F_{\nu_1}(\bm{k},\omega+\Omega_1) F_{\nu_2}(\bm{k},\omega+\Omega_2)\cdots \right]^<$. For one Green's function we have 
\begin{align}
    \int \frac{d\omega}{2\pi}
        \mathrm{tr}\left[
            \sigma_a G^0(\bm{k},\omega)
        \right]^{<}
    &
    = \sum_{\nu_1}
        \mathrm{tr}\left[
            \sigma_a b_{\nu_1}({\bm{k}})
        \right]
        \int \frac{d\omega}{2\pi}
            F^{<}_{\nu_1}(\bm{k},\omega) 
    \nonumber\\
    &= i\sum_{\nu_1}
        \mathrm{tr}\left[
            \sigma_a b_{\nu_1}({\bm{k}})
        \right]
        f(E_{\nu_1}(\bm{k})). 
\end{align}
For two Green's functions we have 
\begin{align}
    &\int \frac{d\omega}{2\pi}
        \mathrm{tr}\left[
            \sigma_a G^0(\bm{k},\omega+\Omega_1) \sigma_b G^0(\bm{k},\omega)
        \right]^{<}
    \nonumber\\
    &= \! \sum_{\nu_1,\nu_2}
        \mathrm{tr}\left[
            \sigma_a b_{\nu_1}({\bm{k}})
            \sigma_b b_{\nu_2}({\bm{k}})
        \right]
        \! \int \! \frac{d\omega}{2\pi}
        \left[
            F_{\nu_1}(\bm{k},\omega \!+\! \Omega_1) F_{\nu_2}(\bm{k},\omega)
        \right]^{<}
    \nonumber\\
    &= -i\sum_{\nu_1,\nu_2}
        \mathrm{tr}\left[
            \sigma_a b_{\nu_1}({\bm{k}})
            \sigma_b b_{\nu_2}({\bm{k}})
        \right]
        \frac{f(E_{\nu_1}(\bm{k}))-f(E_{\nu_2}(\bm{k}))}{\Omega_1^+-\omega_{\nu_1\nu_2}(\bm{k})}, 
\end{align}
where $\Omega_1^+ \equiv \Omega_1 + i 0^+$ and $\omega_{\nu\nu'}(\bm{k})\equiv\omega_{\nu}(\bm{k})-\omega_{\nu'}(\bm{k})$. 
For three Green's functions we have 
\begin{widetext}
\begin{align}
    &\int \frac{d\omega}{2\pi}
        \mathrm{tr}\left[
            \sigma_a G^0(\bm{k},\omega+\Omega_1+\Omega_2)
            \sigma_b G^0(\bm{k},\omega+\Omega_2) \sigma_c G^0(\bm{k},\omega)
        \right]^{<}
    \nonumber\\
    &= \sum_{\nu_1,\nu_2,\nu_3}
        \mathrm{tr}\left[
            \sigma_a b_{\nu_1}({\bm{k}})
            \sigma_b b_{\nu_2}({\bm{k}})
            \sigma_c b_{\nu_3}({\bm{k}})
        \right]
        \int \frac{d\omega}{2\pi}
        \left[
            F_{\nu_1}(\bm{k},\omega+\Omega_1+\Omega_2)
            F_{\nu_2}(\bm{k},\omega+\Omega_2)
            F_{\nu_3}(\bm{k},\omega)
        \right]^{<}
    \nonumber\\
    &= i\sum_{\nu_1,\nu_2,\nu_3}
        \mathrm{tr}\left[
            \sigma_a b_{\nu_1}({\bm{k}})
            \sigma_b b_{\nu_2}({\bm{k}})
            \sigma_c b_{\nu_3}({\bm{k}})
        \right]
        \frac{1}{
            \Omega_1^++\Omega_2^+-\omega_{\nu_1\nu_3}(\bm{k})
        }
        \left[
            \frac{f(E_{\nu_1}(\bm{k}))-f(E_{\nu_2}(\bm{k}))}{
                    \Omega_1^+-\omega_{\nu_1\nu_2}(\bm{k})
                }
            - \frac{f(E_{\nu_2}(\bm{k}))-f(E_{\nu_3}(\bm{k}))}{
                    \Omega_2^+-\omega_{\nu_2\nu_3}(\bm{k})
                }
        \right].
\label{eq:threeGinApp}
\end{align}
\end{widetext}
In the same way we can integrate over $\omega$ in products of $n>3$ Green's functions. 
In the same way we can integrate products of $n>3$ Green's functions with respect to $\omega$. 
In our actual numerical calculations we introduce a finite damping factor $\eta$ by replacing each frequency $\hbar\Omega$ with $ \hbar\Omega + i\eta$ [e.g., for $\Omega_1 = m\Omega$, $\hbar \Omega_1^+  \rightarrow m (\hbar \Omega + i \eta)$], which may correspond to the scheme considering the adiabatic switching of the external field~\cite{passos2018,parker2019,holder2020}.


\section{THG susceptibility} \label{App:THGsus}

Here we summarize the THG susceptibilities corresponding to the diagrams in Fig.~\ref{fig:diagJ}. 
The bare THG  susceptibilities are given by 
\begingroup\allowdisplaybreaks[1]
\begin{widetext}
\begin{align}
    \chi^{0;(3,0)}_{\mu;\mu_1\mu_2\mu_3}(3\Omega;\Omega)
        &= \frac{1}{6} i \left(\frac{e}{\hbar}\right)^4
            \int \! \frac{d\bm{k}}{(2\pi)^d} \! \int \! \frac{d\omega}{2\pi}
            \mathrm{tr}\left[
                \sigma_z G^0(\bm{k},\omega)
            \right]^<
            \xi_{\mu\mu_1\mu_2\mu_3}(\bm{k}), 
    \\
    \chi^{0;(3,1)}_{\mu;\mu_1\mu_2\mu_3}(3\Omega;\Omega)
        &= \frac{1}{2} i \left(\frac{e}{\hbar}\right)^4
            \frac{1}{\hbar} \int \! \frac{d\bm{k}}{(2\pi)^d} \! \int \! \frac{d\omega}{2\pi}
            \mathrm{tr}\left[
                \sigma_z G^0(\bm{k},\omega\!+\!\Omega)
                \sigma_z G^0(\bm{k},\omega)
            \right]^<
            \xi_{\mu\mu_1\mu_2}(\bm{k})
            \xi_{\mu_3}(\bm{k}), 
    \\
    \chi^{0;(3,2a)}_{\mu;\mu_1\mu_2\mu_3}(3\Omega;\Omega)
        &= i \left(\frac{e}{\hbar}\right)^4
            \frac{1}{\hbar^2} \int \! \frac{d\bm{k}}{(2\pi)^d} \! \int \! \frac{d\omega}{2\pi}
            \mathrm{tr}\left[
                \sigma_z G^0(\bm{k},\omega\!+\!2\Omega)
                \sigma_z G^0(\bm{k},\omega\!+\!\Omega)
                \sigma_z G^0(\bm{k},\omega)
            \right]^<
            \xi_{\mu\mu_1}(\bm{k})
            \xi_{\mu_2}(\bm{k})
            \xi_{\mu_3}(\bm{k}), 
    \\
    \chi^{0;(3,2b)}_{\mu;\mu_1\mu_2\mu_3}(3\Omega;\Omega)
        &= \frac{1}{2} i \left(\frac{e}{\hbar}\right)^4
            \frac{1}{\hbar} \int \! \frac{d\bm{k}}{(2\pi)^d} \! \int \! \frac{d\omega}{2\pi}
            \mathrm{tr}\left[
                \sigma_z G^0(\bm{k},\omega\!+\!2\Omega)
                \sigma_z G^0(\bm{k},\omega)
            \right]^<
            \xi_{\mu\mu_1}(\bm{k})
            \xi_{\mu_2\mu_3}(\bm{k}), 
    \\
    \chi^{0;(3,3a)}_{\mu;\mu_1\mu_2\mu_3}(3\Omega;\Omega)
        &=  i \left(\frac{e}{\hbar}\right)^4
            \frac{1}{\hbar^3} \int \! \frac{d\bm{k}}{(2\pi)^d} \! \int \! \frac{d\omega}{2\pi}
            \mathrm{tr}\left[
                \sigma_z G^0(\bm{k},\omega\!+\!3\Omega)
                \sigma_z G^0(\bm{k},\omega\!+\!2\Omega)
                \sigma_z G^0(\bm{k},\omega\!+\!\Omega)
                \sigma_z G^0(\bm{k},\omega)
            \right]^<
            \nonumber\\*&\hspace{270pt}\times
            \xi_{\mu}(\bm{k})
            \xi_{\mu_1}(\bm{k})
            \xi_{\mu_2}(\bm{k})
            \xi_{\mu_3}(\bm{k}), 
    \\
    \chi^{0;(3,3b)}_{\mu;\mu_1\mu_2\mu_3}(3\Omega;\Omega)
        &= \frac{1}{2} i \left(\frac{e}{\hbar}\right)^4
            \frac{1}{\hbar^2} \int \! \frac{d\bm{k}}{(2\pi)^d} \! \int \! \frac{d\omega}{2\pi}
            \mathrm{tr}\left[
                \sigma_z G^0(\bm{k},\omega\!+\!3\Omega)
                \sigma_z G^0(\bm{k},\omega\!+\!2\Omega)
                \sigma_z G^0(\bm{k},\omega)
            \right]^<
            \xi_{\mu}(\bm{k})
            \xi_{\mu_1}(\bm{k})
            \xi_{\mu_2\mu_3}(\bm{k}),                 
    \\
    \chi^{0;(3,3c)}_{\mu;\mu_1\mu_2\mu_3}(3\Omega;\Omega)
        &= \frac{1}{2} i \left(\frac{e}{\hbar}\right)^4
            \frac{1}{\hbar^2} \int \! \frac{d\bm{k}}{(2\pi)^d} \! \int \! \frac{d\omega}{2\pi}
            \mathrm{tr}\left[
                \sigma_z G^0(\bm{k},\omega\!+\!3\Omega)
                \sigma_z G^0(\bm{k},\omega\!+\!\Omega)
                \sigma_z G^0(\bm{k},\omega)
            \right]^<
            \xi_{\mu}(\bm{k})
            \xi_{\mu_1\mu_2}(\bm{k})
            \xi_{\mu_3}(\bm{k}),         
    \\
    \chi^{0;(3,3d)}_{\mu;\mu_1\mu_2\mu_3}(3\Omega;\Omega)
        &= \frac{1}{6} i \left(\frac{e}{\hbar}\right)^4
            \frac{1}{\hbar} \int \! \frac{d\bm{k}}{(2\pi)^d} \! \int \! \frac{d\omega}{2\pi}
            \mathrm{tr}\left[
                \sigma_z G^0(\bm{k},\omega\!+\!3\Omega)
                \sigma_z G^0(\bm{k},\omega)
            \right]^<
            \xi_{\mu}(\bm{k})
            \xi_{\mu_1\mu_2\mu_3}(\bm{k}), 
\end{align}
and the vertex correction terms are given by 
\begin{align}
    \chi^{{\rm vc};(3,2b)}_{\mu;\mu_1\mu_2\mu_3}(3\Omega;\Omega)
        &=- \frac{U}{2}i  \left(\frac{e}{\hbar}\right)^4
              \frac{1}{\hbar} \int \! \frac{d\bm{k}}{(2\pi)^d} \! \int \! \frac{d\omega}{2\pi}\sum_a
            {\mathrm{tr}}\left[
                \sigma_z  G^{0}(\bm{k},\omega\!+\!2\Omega)   \sigma_aG^{0} (\bm{k},\omega)
            \right]^{<}             
              \xi_{\mu\mu_1}(\bm{k}) \Gamma_{\mu_2\mu_3}^a(2\Omega;\Omega,\Omega),
\\
    \chi^{{\rm vc};(3,3b)}_{\mu;\mu_1\mu_2\mu_3}(3\Omega;\Omega)
        &= -\frac{U}{2}i  \left(\frac{e}{\hbar}\right)^4
              \frac{1}{\hbar^2} \int \! \frac{d\bm{k}}{(2\pi)^d} \! \int \! \frac{d\omega}{2\pi}\sum_a
            {\mathrm{tr}}\left[
                \sigma_z  G^{0}(\bm{k},\omega\!+\!3\Omega)  \sigma_zG^{0} (\bm{k},\omega\!+\!2\Omega) \sigma_a G^{0} (\bm{k},\omega)
            \right]^{<}
            \nonumber\\* &\hspace{235pt}\times
                         \xi_{\mu}(\bm{k})\xi_{\mu_1}(\bm{k})\Gamma_{\mu_2\mu_3}^a(2\Omega;\Omega,\Omega),     
                        \\                  
    \chi^{{\rm vc};(3,3c)}_{\mu;\mu_1\mu_2\mu_3}(3\Omega;\Omega)
        &= -\frac{U}{2}i  \left(\frac{e}{\hbar}\right)^4
              \frac{1}{\hbar^2} \int \! \frac{d\bm{k}}{(2\pi)^d} \! \int \! \frac{d\omega}{2\pi}\sum_a
            {\mathrm{tr}}\left[
                \sigma_z  G^{0}(\bm{k},\omega\!+\!3\Omega)  \sigma_aG^{0} (\bm{k},\omega\!+\!\Omega) \sigma_z G^{0} (\bm{k},\omega)
            \right]^{<}
                        \nonumber\\*&\hspace{235pt}\times
                                     \xi_{\mu}(\bm{k})\Gamma_{\mu_1\mu_2}^a(2\Omega;\Omega,\Omega)\xi_{\mu_3}(\bm{k}).                                                              
\end{align}
\end{widetext}
\endgroup

\begin{figure}[h]
    \begin{minipage}[t]{0.8\columnwidth}
        \begin{center}
        \includegraphics[bb=0 20 460.8 345.6,width=1\columnwidth]{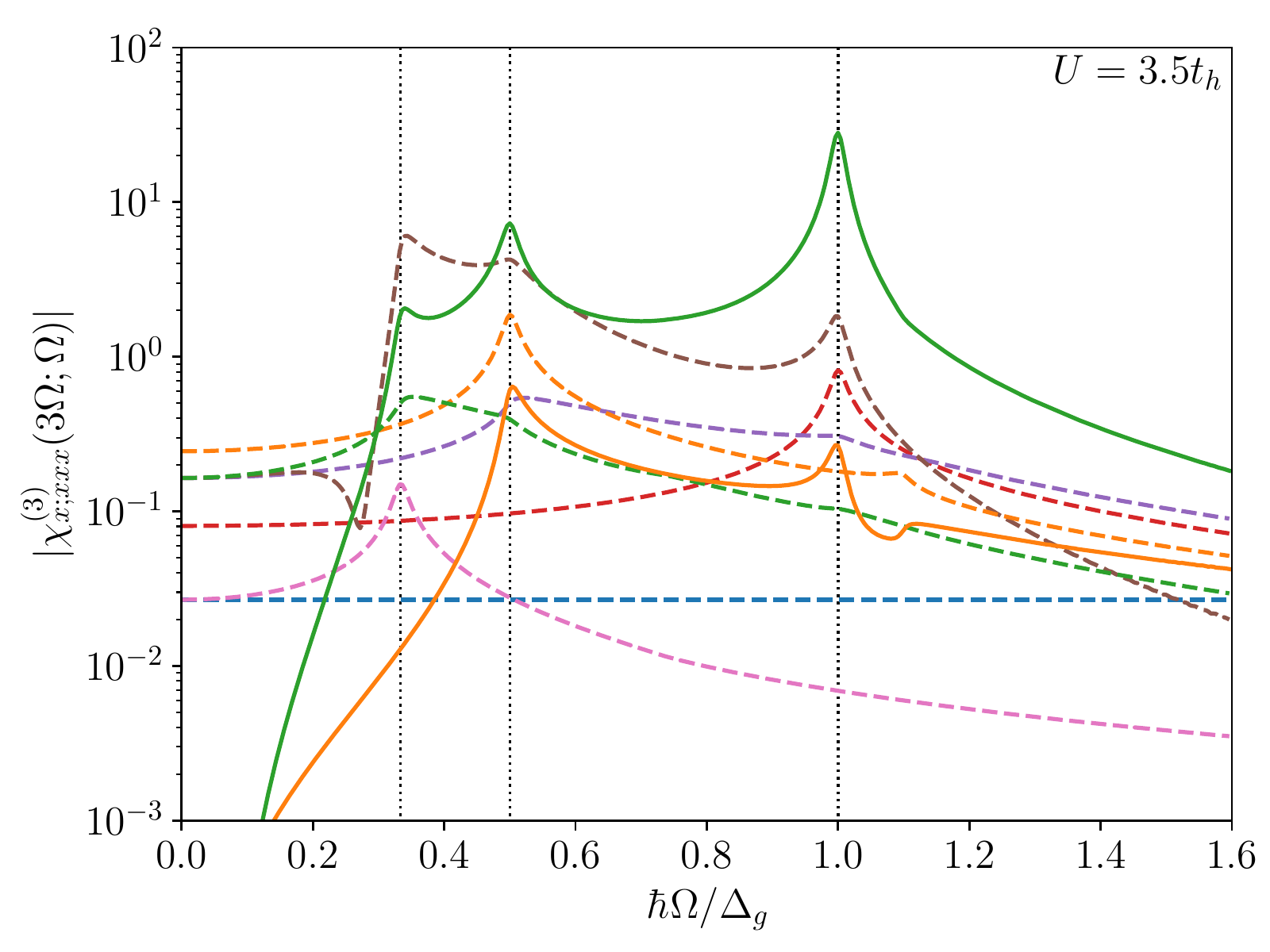}
        \end{center}
    \end{minipage}%
    \begin{minipage}[t]{0.2\columnwidth}
        \begin{center}
        \includegraphics[bb=0 10 130 360,width=1\columnwidth]{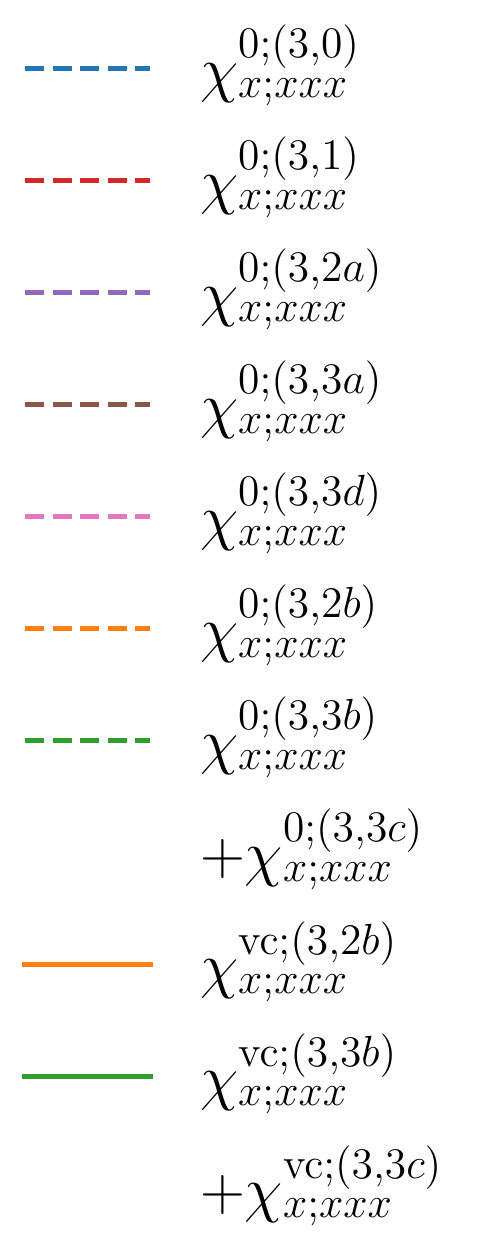}
        \end{center}
    \end{minipage}
    \caption{THG susceptibility  decomposed into the bare susceptibilities and vertex corrections, where $U =3.5 t_h$, $D = \Delta_0 - \Delta_1 = 3.8 t_h$, and $\eta= 0.01\Delta_g$ are assumed. The vertical dotted lines indicate $\hbar \Omega = \Delta_g/3$, $\Delta_g/2$, and $\Delta_g$.}
    \label{fig:suscep_omega_term}
\end{figure}

In Fig.~\ref{fig:suscep_omega_term} we present all components of the bare THG susceptibility $\chi^{0;(3)}_{x;xxx}(3\Omega;\Omega)$ and vertex correction $\chi^{{\rm vc};(3)}_{x;xxx}(3\Omega;\Omega)$.  
Among the bare susceptibilities, 
the component 3a is the largest and mainly contributes to THG at $\hbar \Omega = \Delta_g/3$. 
Corresponding to Fig.~\ref{fig:suscep_omega}(e), the vertex correction 3b$+$3c is the largest at  $\hbar \Omega = \Delta_g/2$ and $\Delta_g$. 

\begin{figure}[b]
    \begin{center}
    \includegraphics[bb=0 40 460.8 347,width=0.88\columnwidth]{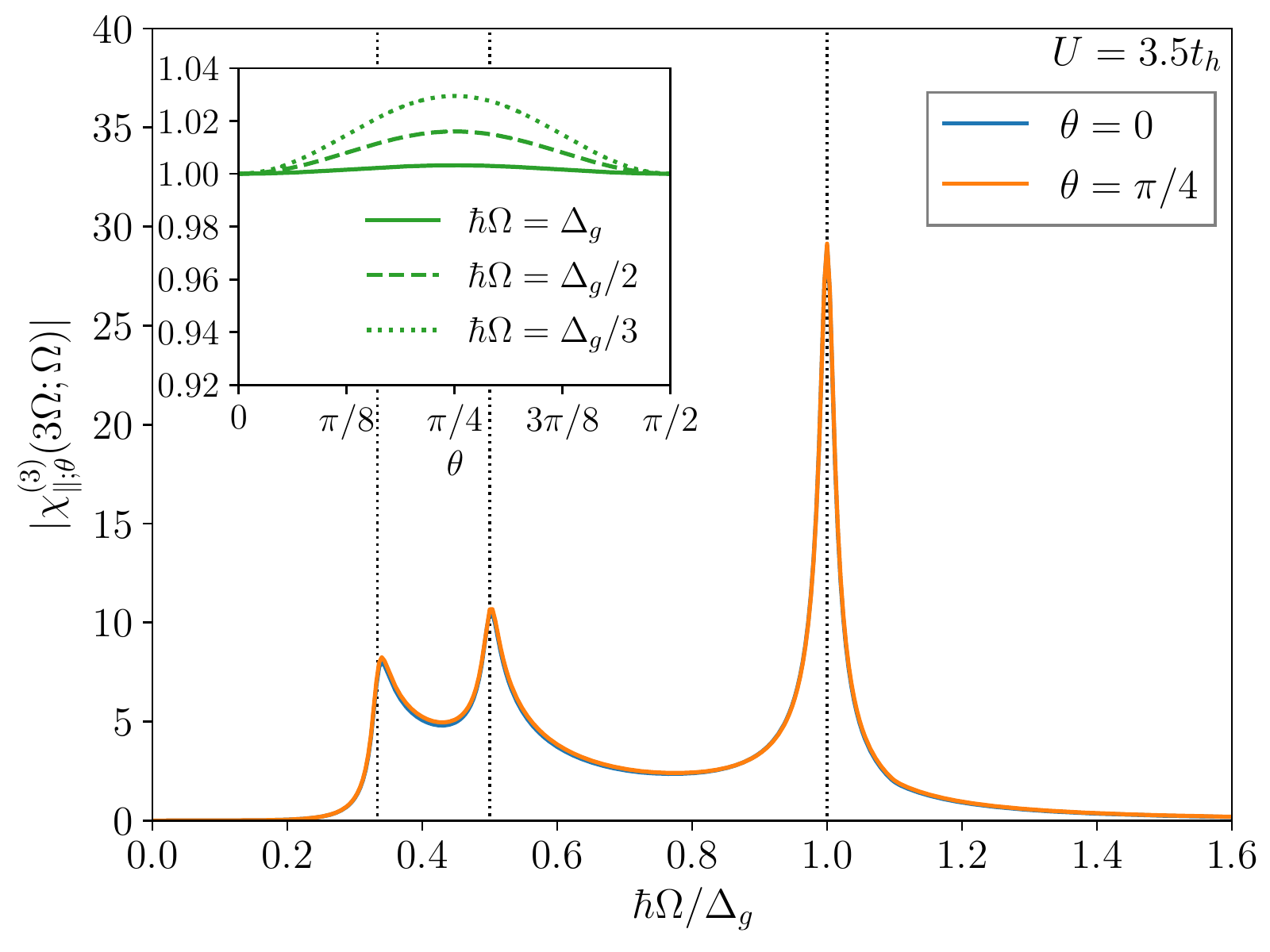}
    \end{center}
    \caption{
    THG susceptibility $|\chi^{(3)}_{\parallel,\theta}(3\Omega;\Omega)|$ at $\theta=0$ and $\pi/4$, 
     where $U =3.5 t_h$, $D = \Delta_0 - \Delta_1 = 3.8 t_h$, and $\eta= 0.01\Delta_g$ are used. The vertical dotted lines indicate $\hbar \Omega = \Delta_g/3$, $\Delta_g/2$, and $\Delta_g$. 
    Inset: Polarization dependence of the normalized $|\chi^{(3)}_{\parallel,\theta}(3\Omega;\Omega)|$ at $\hbar \Omega = \Delta_g/3$, $\Delta_g/2$, and $\Delta_g$, where the susceptibilities are normalized to their values at $\theta=0$. 
    }
    \label{fig:suscep_polarization}
\end{figure}


\section{Polarization dependence} \label{App:pol}

Here we show the polarization dependence of the THG susceptibility. 
When the external field 
\begin{align}
    \bm{A}(\Omega) = A(\Omega)(\cos\theta\,\bm{e}_x+\sin\theta\,\bm{e}_y) = A(\Omega)\hat{\bm{n}}^{(\theta)} 
\end{align}
 is applied, the THG susceptibility parallel to the polarization direction $\hat{\bm{n}}^{(\theta)}$ is given by 

\begin{align}
    \chi^{(3)}_{\parallel,\theta}(3\Omega;\Omega)
    = \sum_{\mu,\mu_1,\mu_2,\mu_3}
        \chi^{(3)}_{\mu;\mu_1\mu_2\mu_3}(3\Omega;\Omega)
        n_{\mu}^{(\theta)}n_{\mu_1}^{(\theta)}n_{\mu_2}^{(\theta)}n_{\mu_3}^{(\theta)}, 
\end{align}
where $\theta$ is the angle with respect to the $x$ axis, and $n_x^{(\theta)}=\cos\theta$ and $n_y^{(\theta)}=\sin\theta$. 

Figure~\ref{fig:suscep_polarization} shows the polarization dependence of the THG susceptibility $|\chi^{(3)}_{\parallel,\theta}(3\Omega;\Omega)|$.
Even when the incident light is polarized along the $\theta = \pi/4$ direction, $\chi^{(3)}_{\parallel,\theta}(3\Omega;\Omega)$ retains the main features of the THG susceptibility observed at $\theta = 0$. 
The difference in $|\chi^{(3)}_{\parallel,\theta}(3\Omega;\Omega)|$ at $\hbar \Omega = \Delta_g/3$ is less than 4\% and the others are smaller than that (see the inset of Fig.~\ref{fig:suscep_polarization}). 
In particular, $|\chi^{(3)}_{\parallel,\theta}(3\Omega;\Omega)|$ at $\hbar \Omega = \Delta_g$ is almost flat with respect to $\theta$. 
Therefore, the polarization dependence of THG is small in the EI.


\bibliography{References}

\begin{thebibliography}{99}
\bibitem [{fin()}]{finiteT}%
  \BibitemOpen
  \href@noop {} {}\bibinfo {note} {While we assume the order parameter at
  finite temperature in the 2D model with the continuous symmetry, the ordered
  states in the quasi-2D EI candidates may be characterized by the broken
  discrete symmetries due to additional factors (e.g. electron-phonon
  coupling)~\cite{zenker2014,kaneko2015}, so that we expect that the tendencies
  of our main results are comparable with THG in the candidate
  materials.}\BibitemShut {Stop}%
\end{thebibliography}

\end{document}